\documentclass[english]{article}
\usepackage[T1]{fontenc}
\usepackage[utf8]{inputenc}
\usepackage{geometry}
\usepackage{babel}
\usepackage{float}
\usepackage{textcomp}
\usepackage{amsmath}
\usepackage{amsthm}
\usepackage{graphicx}
\usepackage{setspace}
\usepackage[unicode=true,pdfusetitle,
 bookmarks=true,bookmarksnumbered=false,bookmarksopen=false,
 breaklinks=false,pdfborder={0 0 1},backref=false,colorlinks=false]
 {hyperref}

\makeatletter
\theoremstyle{plain}
\newtheorem{lyxalgorithm}{\protect\algorithmname}
\theoremstyle{plain}
\newtheorem{criterion}{\protect\criterionname}
\theoremstyle{definition}
\newtheorem{condition}{\protect\conditionname}
\theoremstyle{definition}
 \newtheorem{example}{\protect\examplename}

\makeatother

\providecommand{\algorithmname}{Algorithm}
\providecommand{\conditionname}{Condition}
\providecommand{\criterionname}{Criterion}
\providecommand{\examplename}{Example}

\begin{document}
\begin{center}
\textbf{\LARGE{}DeFi Security: Turning The Weakest Link Into The Strongest
Attraction}{\LARGE\par}
\par\end{center}

\begin{center}
\textbf{\large{}Ravi Kashyap (ravi.kashyap@stern.nyu.edu)}\footnote{\begin{doublespace}
Numerous seminar participants, particularly at a few meetings of the
econometric society and various finance organizations, provided suggestions
to improve the paper. The following individuals have been a constant
source of inputs and encouragement: Dr. Yong Wang, Dr. Isabel Yan,
Dr. Vikas Kakkar, Dr. Fred Kwan, Dr. Costel Daniel Andonie, Dr. Guangwu
Liu, Dr. Jeff Hong, Dr. Humphrey Tung and Dr. Xu Han at the City University
of Hong Kong. The views and opinions expressed in this article, along
with any mistakes, are mine alone and do not necessarily reflect the
official policy or position of either of my affiliations or any other
agency.
\end{doublespace}
}
\par\end{center}

\begin{center}
\textbf{\large{}Estonian Business School / City University of Hong
Kong}{\large\par}
\par\end{center}

\begin{center}
\today
\par\end{center}

\begin{center}
\par\end{center}

\begin{onehalfspace}
\begin{center}
Association for Computing Machinery Classification System: K.6.5:
Security and Protection; D.4.6: Security and Protection; D.2.11: Software
Architectures; B.8.2: Performance Analysis and Design Aids; I.2.8:
Problem Solving; H.1.1: Systems and Information Theory
\par\end{center}
\end{onehalfspace}

\begin{doublespace}
\begin{center}
Journal of Economic Literature Codes: O3 Innovation • Research and
Development • Technological Change • Intellectual Property Rights;
D81 Criteria for Decision-Making under Risk and Uncertainty; C63 Computational
Techniques; O31 Innovation and Invention: Processes and Incentives 
\par\end{center}

\begin{center}
\par\end{center}
\end{doublespace}

\begin{center}
\tableofcontents{}\pagebreak{}
\par\end{center}

\begin{doublespace}
\begin{center}
\listoffigures 
\par\end{center}

\begin{center}
\listoftables 
\par\end{center}
\end{doublespace}

\begin{singlespace}
\begin{center}
\pagebreak{}
\par\end{center}
\end{singlespace}
\begin{doublespace}

\section{Abstract}
\end{doublespace}

The primary innovation we pioneer - focused on blockchain information
security- is called the Safe-House. The Safe-House is badly needed
since there are many ongoing hacks and security concerns in the DeFi
space right now. The Safe-House is a piece of engineering sophistication
that utilizes existing blockchain principles to bring about greater
security when customer assets are moved around. The Safe-House logic
is easily implemented as smart contracts on any decentralized system.
The amount of funds at risk from both internal and external parties
- and hence the maximum one time loss - is guaranteed to stay within
the specified limits based on cryptographic fundamentals. 

The Safe-House provides a first level of security that caters to,
a significant extent, the enhanced fund movement needs for today's
DeFi protocols and blockchain investment funds. To improve the safety
of the Safe-House even further, we adapt the one time password (OPT)
concept to operate using blockchain technology. Well suited to blockchain
cryptographic nuances, our secondary advancement can be termed the
one time next time password (OTNTP) mechanism. The OTNTP is designed
to complement the Safe-House making it even more safe. 

We provide a detailed threat assessment model - discussing the risks
faced by DeFi protocols and the specific risks that apply to blockchain
fund management - and give technical arguments regarding how these
threats can be overcome in a robust manner. We discuss how the Safe-House
can participate with other external yield generation protocols in
a secure way. We provide reasons for why the Safe-House increases
safety without sacrificing the efficiency of operation. We start with
a high level intuitive description of the landscape, the corresponding
problems and our solutions. We then supplement this overview with
detailed discussions including the corresponding mathematical formulations
and pointers for technological implementation.This approach ensures
that the article is accessible to a broad audience. 

Investors are currently wary of putting their trust - and wealth -
in the hands of financial intermediaries when they cannot be held
fully accountable in an industry that still requires much closer scrutiny
and regulation. Our innovations will ensure that losses can be attributed
- mostly - to oversights - or any fraudulent intentions - of fund
representatives who are accepting money from others and hence are
taking on fiduciary responsibility. Such accountability - and transparency
- will bring greater interest - and involvement - from many who are
still on the fringes - or even far away from the reach - of blockchain
technology. Incorporating the best practices we have introduced should
aid in the wider adoption of blockchain technologies and in particular
assuage the concerns of those who are disturbed by the increasing
number of crimes in this space. 

When investors are assured about the safety of their funds, it will
provide much needed impetus - in terms of financing, by allowing many
superior risk mitigation techniques from traditional finance to be
applied to blockchain portfolios - to several upcoming projects by
enhancing trust between financial market participants and intermediaries
operating in a decentralized environment. We have reviewed around
240 papers spanning both technology and financial topics - artificial
as these topical distinctions can be at times, we have strived to
ensure that the paper can be useful for both technologists and financial
specialists - and linked the insights towards creating a secure wealth
management system that can be accessible by everyone. 

The Safe-House - and related components - have been successfully deployed
for commercial use on multiple platforms - Ethereum, Binance and Polygon
- and hundreds of transactions have been performed satisfactorily. 

\section{\label{sec:Introduction:-DeFi-Defense}Introduction: DeFi Defense
Dilemmas}

The invention of Bitcoin is transforming all aspects of business and
human interactions (Nakamoto 2008; Narayanan \& Clark 2017; Chen 2018;
Monrat, et al. 2019; End-note \ref{enu:Blockchain-Terms}). This seminal
event - a landmark permanently etched in the history of technological
change - has opened the floodgates for innovations seeking to add
different aspects of human experiences onto the blockchain (Karamitsos
et al., 2018; Zhang et al., 2018; Alladi et al., 2019; Kuo et al.,
2019; Lu 2019; Whitaker 2019; Bumblauskas et al., 2020; Gatteschi
et al., 2020; Hunhevicz \& Hall 2020; Prewett et al., 2020; Brophy
2020). The rest, as they say, is history. 

Most blockchain services have a financial component embedded within
them as part of the overall offering. Almost all decentralized projects
- even those which do not directly cater to a financial need - create
blockchain based tokens - which are termed as cryptocurrencies - to
facilitate transactions for their customers. The collection of these
financial services using decentralized ledger technology can be denoted
by the term decentralized finance (DeFi: Zetzsche, Arner \& Buckley
2020; Werner et al., 2021; Grassi et al., 2022; End-note \ref{enu:Decentralized-finance}). 

DeFi applications are bringing many traditional financial utilities
- and wealth generation techniques - onto the decentralized realm.
The rapid growth of this sector is bound to continue since new blockchain
projects will focus on rolling out various products, which become
investment opportunities. As newer blockchain platforms develop -
and seek financing - investors will look to diversify their risks
by spread out their investments across different networks. To continually
monitor the risk and return profiles of such a portfolio spread across
multiple networks - and tweak the holdings to suit risk preferences
based on market conditions - would be an extremely arduous task -
almost impossible for non-sophisticated investors (De Bondt 1998;
Elton et al., 2009; Fuertes et al., 2014; Torre \& Rudd 2004). 

The complexity of managing wealth - and risks - across several platforms
will give rise to specialized blockchain wealth managers. Such blockchain
investment vehicles are actively trying to solve investment problems
- more relevant to blockchain - by adapting many best practices from
traditional finance for the blockchain realm (Cai 2018; Peterson 2018;
Arshadi 2019; Schär 2021; Kareem 2021-II; 2022; Dos Santos et al.,
2022). These blockchain-funds will collect money from several investors
and invest in several assets across different networks. Decentralized
financial services - which are accessible to everyone - providing
sophisticated money management to the masses can now become a reality. 

A tragic trend in this remarkable technological saga is the narrative
of hacks, security vulnerabilities and related exploits - leading
to significant loss of funds for several participants. We can broadly
categorize the fundamental reasons for blockchain vulnerabilities
into three groups: 1) information management issues - related to misuse
of passwords, cryptographic keys and abuse of trust; 2) technical
issues - server or infrastructure breaches, coding logic or smart
contract bugs, cloud platform issues; and 3) economic incentives being
misused or flaws in the financial principles used. 

Chia et al., 2018; Zamani et al., (2020); Guo \& Yu (2022); survey
several blockchain security lapses seeking to identify the root cause
of these issues, which can provide lessons for other participants
to bolster their security. Wang et al., 2021; Werapun et al., (2022)
look at how flash loans can be used to steal funds from DeFi protocols.
Lee et al., 2022; Uhlig (2022); Briola et al., (2023) is an account
of market failure related to the implosion of a DeFi project.

Li et al., (2022); Li et al., (2022) provide a systematic summary
of DeFi security incidents and systematically analyze the vulnerabilities.
Zhou et al., (2022) construct a DeFi reference frame that categorizes
77 academic papers, 30 audit reports, and 181 incidents, which reveals
the differences in how academia and the practitioners’ community defend
and inspect incidents - including both attacks and accidents. Wang
et al., (2022) propose a deep-learning-based attack detection system
to characterize DeFi attacks. 

Chaliasos et al., (2023) investigate the effectiveness of automated
security tools in identifying vulnerabilities that can lead to attacks.
Their findings reveal that automated tools could have prevented a
mere 8\% of the attacks in which amounts to \$149 million out of the
\$2.3 billion in losses across the 127 attacks they evaluated. 

Liu et al., (2022) propose a semi-centralized trust management architcture
that can identify and mitigate malicious influences. Liu et al., (2022)
propose a blockchain-empowered federated learning framework for healthcare-based
cyber physical systems (Pokhrel \& Choi 2020; End-note \ref{enu:Federated-learning-(also}).
Tian et al., (2019) propose a secure digital evidence framework using
blockchain with a loose coupling structure in which the evidence and
the evidence information are maintained separately. Liu et al., (2022)
propose the design of a blockchain-enabled reputation system - in
which the ratings' privacy is strongly preserved in the processes
of transmission and storage - by especially considering the rating
privacy issue. 

Other than the immediate loss of funds, thefts and security incidents
lead to loss of confidence for participants and increase volatility
in the markets. Grobys (2021); Chen et al., (2023) investigate hacking
incidents in the Bitcoin market and find that the volatility increases
significantly. Corbet et al., (2020) find indications that blockchain
crime can result in substantial loss of confidence in the cryptocurrency
market. Moore \& Christin (2013) find that - not surprisingly - the
more successful blockchain projects are likely to attract hackers
and experience a breach. 

Investors are currently wary of putting their trust - and wealth -
in the hands of financial intermediaries when they cannot be held
fully accountable in an industry that still requires much closer scrutiny
and regulation (Guo \& Liang 2016; Yeoh 2017; Girasa 2018; Yeung 2019;
Shanaev et al., 2020; Yadav et al., 2022). Our innovations will ensure
that losses can be attributed - mostly - to oversights - or any fraudulent
intentions - of fund representatives who are accepting money from
others and hence are taking on fiduciary responsibility. Such accountability
- and transparency - will bring greater interest - and involvement
- from many who are still on the fringes - or even far away from the
reach - of blockchain technology.

\subsection{\label{subsec:Enter-The-Safe-House:}Enter The Safe-House: With An
OTNTP (One-Time Next-Time Password)}

A common requirement for all blockchain based wealth management protocols
will be to utilize the principles of investing to accomplish better
risk mitigation. Along with financial risk mitigation, having proper
safety mechanisms in place - to prevent thefts and other security
incidents - will be extremely crucial requirement. As discussed earlier:
several technical enhancements, software architectural designs and
elaborate powerful tools are being created to remedy blockchain related
security issues. Improved economic incentives and risk mechanisms
are being insituted within projects based on an understanding of previous
failures. These intricate improvements will no doubt serve to better
the infrastructure for blockchain transactions. 

What is lacking is a security architecture - and software design -
that caters to the fund movement requirements of blockchain wealth
management protocols. In this paper we discuss what we consider to
be the foremost priority for all organizations engaged in decentralized
finance endeavors. What DeFi badly needs is a strengthened security
blueprint. DeFi Protocols need to add protective shields against internal
theft and external intrusion. The security related features have to
be an intrinsic part of the design of distributed applications and
deeply embedded within the software architecture.

The innovation we pioneer - focused on blockchain security - is called
the Safe-House. The Safe-House is badly needed since there are many
ongoing hacks and security concerns in the DeFi space right now. The
Safe-House is a piece of engineering sophistication that utilizes
existing blockchain principles to bring about greater security when
customer assets are moved around. The Safe-House logic is easily implemented
as smart contracts on any decentralized system (Wang et al., 2018;
Mohanta et al., 2018; Zou et al., 2019; Zheng et al., 2020; End-note
\ref{enu:A-smart-contract}). The amount of funds at risk from both
internal and external parties - and hence the maximum one time loss
- is guaranteed to stay within the specified limits based on cryptographic
fundamentals. 

The Safe-House provides a first level of security that caters to a
significant extent the enhanced fund movement needs for today's DeFi
protocols. To improve upon this further, we adapt the one time password
- (OTP: Haller et al., 1998; Goyal et al., 2005; M'Raihi et al., 2011;
Aravindhan \& Karthiga 2013; Erdem \& Sandıkkaya 2018; End-note \ref{enu:A-one-time-password})
- concept to operate within the constraints of blockchain technology
tightly embedded within the working of the safe-house. Well suited
to blockchain cryptographic nuances, our secondary advancement, termed
the one time next time password (OTNTP) mechanism, is designed to
complement the Safe-House making it even more safe. We provide a detailed
discussion - including the corresponding mathematical formulations
and pointers for technological implementation.

\textbf{\textit{Necessity is the mother of all invention - or creation
/ innovation - but the often forgotten father is frustration.}}

The enhanced security features we describe here are - no doubt - very
necessary. But the essence of the security innovations we are creating
are borne out of the numerous troubles - frustrations - several (all?)
protocols are encountering due to unauthorized parties trying to access
their funds. The same - frustration being the motivation - could be
said about the rest of the investment vehicles discussed in Kasaliya
(2022). These innovations are very necessary. But the key motivation
for these mechanisms - and architectural designs - are due to the
main issues that one encounters while trying to obtain: 1) unencumbered
access to decent investment opportunities in the traditional financial
world; and 2) peace of mind while investing in crypto assets.

We start with a high level intuitive description of the landscape,
the corresponding problems and our solutions. We then supplement this
overview with detailed discussions including the corresponding mathematical
formulations and pointers for technological implementation.This approach
ensures that the article is accessible to a broad audience. We have
reviewed around 240 papers spanning both technology and financial
topics - artificial as these topical distinctions can be at times,
we have strived to ensure that the paper can be useful for both technologists
and financial specialists - and linked the insights towards creating
a secure wealth management system that can be accessible by everyone. 

Incorporating the best practices we have introduced should aid in
the wider adoption of blockchain technologies and in particular assuage
the concerns of those who are disturbed by the increasing number of
crimes in this space. When investors are assured about the safety
of their funds, it will provide much needed impetus - in terms of
financing, by allowing many superior risk mitigation techniques from
traditional finance to be applied to blockchain portfolios - to several
upcoming projects by enhancing trust between financial market participants
and intermediaries operating in a decentralized environment. 

\subsection{\label{subsec:Outline-of-the}Outline of the Sections Arranged Inline}

Section (\ref{sec:Introduction:-DeFi-Defense}) - which we have already
seen - provides an introductory overview emphasizing the importance
of information security in decentralized platforms. It summarizes
our contributions to increasing DeFi security through the design of
a novel software architecture suited for fund movements in the blockchain
realm. Section (\ref{subsec:Enter-The-Safe-House:}) - which we have
also already seen - summarizes the novel contributions we are making
to DeFi wealth management security improvements.

Section (\ref{sec:The-Safe-House}) gives a high level intuitive description
of our innovations to strengthen blockchain security using a novel
paradigm we have created - termed the Safe-House - including a discussion
of how the Safe-House functions. Section (\ref{subsec:Staying-on-TOP})
gives a summary of the One Time Password (OPT) technique that has
become quite prevalent in identity management and fraud prevention.
Section (\ref{subsec:Safe-House-Safety-Without}) gives the rationale
for why the Safe-House increases safety without sacrificing the efficiency
of operation. Section (\ref{subsec:Tall-Towers-Weak-Links}) gives
an intuitive analogy to describe why strengthening security can make
DeFi an extremely attractive proposition for participants and providers
alike.

A detailed threat assessment model - discussing the risks faced by
DeFi protocols and the specific risks that apply to blockchain fund
management - is outlined in Section (\ref{subsec:Threat-Assessment-Model}).
Section (\ref{subsec:Components-of-Blockchain}) describes the main
components of any blockchain fund management system. Section (\ref{subsec:Fund-Loss-Categories})
chronicles the various means through which investment funds - with
many general concepts applicable outside blockchain, but with various
pointers specific to decentralized systems - can face losses and even
result in liquidation. Section (\ref{sec:Centralized-Risk:-Watching})
gives a broad set of different approaches to address blockchain fund
movement security concerns. 

Section (\ref{sec:The-Safe-House-Combination:}) is a discussion of
the specifics related to the Safe-House - with mathematical formulations
and detailed explanations - that can be helpful for actual software
implementations. Here we give detailed technical arguments regarding
how the threats discussed in Section (\ref{subsec:Threat-Assessment-Model})
can be overcome in a robust manner. Section (\ref{subsec:The-Safe-House-Architecture})
gives the architectural diagram for blockchain fund management using
the Safe-House and our other innovations. Section (\ref{subsec:Multiple-Signatures,-One-Time-Ne})
gives the detailed - step by step - methodology that brings greater
security to decentralized investing. Section (\ref{subsec:The-One-Time-Next-Time-Password-})
is a formal discussion of the one time next time password (OTNTP)
concept that we have advanced to complement the Safe-House. Section
(\ref{subsec:Yield-Enhancement-Through}) considers the problem of
transferring funds to external yield enhancement protocols and our
solution to securely accomplish this goal. 

Sections (\ref{sec:Areas-for-Further}; \ref{sec:Conclusion}) suggest
further avenues for improvement and the conclusions respectively.
Appendix (\ref{sec:End-notes}) in Part (\ref{part:Appendix-of-Supplementary})
has supplementary explanations for the concepts mentioned in the main
text.

\section{\label{sec:The-Safe-House}The Safe-House: DeFinitely Strengthens
DeFi Security Infinitely}

In today’s blockchain environment, many protocols are constantly under
threat wherein their assets can be taken out or withdrawn by unlicensed
external actors. The problem is compounded since internal parties
that can access organizational technology assets - for fund management
and operational purposes or software development and maintenance -
might try to steal funds that are not meant for their use or utilize
blockchain resources to devise schemes to dupe investors (Bartoletti
et al., 2020; Grobys 2021; Trozze, Kleinberg \& Davies 2021; Li et
al., 2022; Trozze et al., 2022; Kaur et al., 2023; Wang et al., 2021).
Cryptographic methods used in blockchain protocols do provide a certain
amount of security. But most projects are still vulnerable either
when cryptographic keys - corresponding to fund movements - are compromised
or when internal parties - who have access to the keys - have the
intention of misappropriating investor funds.

The extent of the perils are magnified in the blockchain environment
- since a few parties with malicious intent can reach numerous victims
- given the distributed nature of this technology. This adds to the
perception that security dangers are commonplace and that hackers
are ruling the roost. The many security related incidents stand in
the way of the mass adoption of blockchain technology, which otherwise
has the potential to transform all human interactions. We wish to
do our part to grow this ecosystem by mitigating the harmful influences
and restoring the balance of power to groups that are actively trying
to develop this landscape.

To counter these hazards, we are introducing several new innovations
that will increase the overall defense mechanisms of any blockchain
fund management protocol. The novel security innovations - which we
are developing - are to ensure that a DeFi investment fund cannot
be compromised by either internal or external actors. Our multi-pronged
protection scheme refines the existing cryptographic cover by adding
extra layers of protective shields. By making these upgrades, we are
converting one of the major drawbacks of the DeFi space to one of
the major strengths of any protocol.

The central element of our security innovations is the creation of
a safe house - which will be guarded by private-public key cryptographic
methods (Bernstein \& Lange 2017; Kaushik et al., 2017; Puthal et
al., 2018; Stephen \& Alex 2018; McBee \& Wilcox 2020; Pal et al.,
2021; End-note \ref{enu:Public-key-cryptography,-or}) - to store
all blockchain financial assets. As an additional measure to enhance
the security, access to the safe house will be provided only upon
verification of the identity of the person requesting the permission.
Our identity verification methodology is above and beyond the security
provided by existing blockchain public-private key cryptographic methods.

We call our authentication technique the One Time Next Time Password
(OTNTP). The OTNTP concept will be used to verify the identity of
the portfolio manager - trying to take out funds - and to allow Safe-House
access for making withdrawals. Our modified authentication mechanism
- OTNTP - should help with password protection in decentralized environments
where all transaction information has to be made public for verification
purposes. 

\subsection{\label{subsec:Staying-on-TOP}Staying on TOP of The OTP (One Time
Password)}

The OTNTP - the novel technique we advance for blockchain systems
- is a modification of the One Time Password (OTP) mechanism. The
OTP approach that has become quite prevalent in identity management
and fraud prevention. As the name suggests - One Time Passwords -
are valid only once. Aravindhan \& Karthiga (2013) provide a survey
of the OTP mechanism. Using OTPs provides a greater level of security
to information systems - that require passwords to gain access - since
a new password is needed each time. Information systems that have
static passwords - that do not change for prolonged periods - are
compromised - guessed or stolen - by increasingly sophisticated hackers. 

Originating with the proposal made by Lamport (1981), OTPs have found
many use cases for financial activity and identity verification purposes.
Several improvements and refinements have been made to the OTP concept.
Groza \& Petrica (2005) suggest the use of functions on groups of
composite integers instead of one way hash functions (Naor \& Yung
1989; Merkle 1990; End-note \ref{enu:One-Way-Function}; \ref{enu:A-one-way-hash}).
Goyal et al., (2005) suggest a way to reduce client computational
requirements - associated with the OTP - without a significant increase
in the server computational requirements. M'Raihi et al., (2011) outline
a time-based variant of the OTP algorithm that provides short-lived
OTP values, which are desirable for enhanced security. Erdem \& Sandıkkaya
(2018) describe an architecture to outsource the OTP verification
to cloud computing platforms, which should ease adoption usage both
for users and service providers (Ray 2017; End-note \ref{enu:Cloud-computing}).

\subsection{\label{subsec:Safe-House-Safety-Without}Safe-House Safety Without
Sacrificing Efficiency}

The safe house has also been designed to detect and neutralize dangers
such as attempts to withdraw by players without the right credentials.
If a real threat is determined, the safe house will go into a locked
state. It will not allow anyone to take out any assets or funds from
it until the severity of the danger has been assessed and it is deemed
safe to resume further operations. It would be helpful to keep a detailed
history of all the transactions - facilitated by decentralized ledger
technology - linked to specific internal staff responsible for fund
movements and trade execution. Needless to say, an extra layer of
protection will be provided if the personnel involved in the process
are fully KYC’ed (Know Your Customer or Client, or in this case KYE'ed,
Know Your Employees: Bilali 2011; Chen 2020; Ostern \& Riedel 2021;
Malhotra, Saini \& Singh 2022; End-note \ref{enu:KYC}). 

In the event of an extreme situation - such as a malicious party breaching
the safe house - the extent of damage will be limited due to numerous
safeguards on the mobility of funds. This scenario can occur if an
internal member - or an employee - decides to turn rogue. In such
a case, the identity of the person who stole the funds would be established
with certainty, due to the incorporation of appropriate identity verification
methodologies and the amount lost would be minimal. Even if the missing
amount is very small, further action can be taken to recover the lost
funds since the identity of the individual - who took the funds -
will be known.

While building the new security features mentioned above, the overriding
challenge will be to ensure that the improved safety procedures will
not become too cumbersome. The objective is to be able to accommodate
more security guidelines and yet operate quickly - and effectively
- to take advantage of market conditions. This can be accomplished
by matching fund flows - which are governed by Safe-House security
parameters - to asset management principles and requirements. A detailed
discussion of our trade execution related innovations given in Kareem
(2021-I) demonstrate that funds movements have to happen in batches
- which requires large fund movement requirements to be split into
many smaller transfers - since that is the way to minimize trading
transaction costs and related issues. To summarize, the result is
a system that will protect investor assets and yet allow smooth functioning
of the investment machinery. 

\subsection{\label{subsec:Tall-Towers-Weak-Links}Tall Towers and Weak Links}

We give an intuitive analogy to describe why strengthening security
can make DeFi an extremely attractive proposition for participants
and providers alike. This justification appears in the title of the
paper.

Any tall tower has to withstand a lot of wind resistance (Nagase,
Hisatoku \& Yamazaki 1993; Simiu \& Scanlan 1996; Deng et al., 2019;
Gu 2010; Li et al., 2020). The taller a structure the stronger the
wind forces that it has to overcome. Hence, the height of the tall
tower becomes its weakest aspect. But if this weakness is addressed
properly, and enough safety mechanisms are incorporated in the design,
the height of the tower becomes its greatest attraction. People flock
to the top to marvel at the views and to admire the accomplishment
of having created such a safe and tall structure. Clearly, the importance
of having a solid foundation for a tall structure cannot be overlooked.

Likewise, security is the biggest threat - or the weakest link - in
DeFi right now. DeFi is nothing but the movement of funds seeking
profits. The more the funds move, the greater the security vulnerability.
But if the security concerns are adequately addressed - and appropriate
features are designed to make DeFi investing more safe - this very
weakness can be turned into the greatest attraction. The captivating
fascination will then be the generation of significant wealth for
all participants. The solid foundation - in our case - is the rigorous
risk management that has to be an intrinsic part of the DeFi framework
(Kasaliya 2022).

\section{\label{subsec:Threat-Assessment-Model}Threat Assessment Model for
DeFi Wealth Management}

\subsection{\label{subsec:Components-of-Blockchain}Components of Blockchain
Wealth Management Platforms}

\begin{doublespace}
\noindent 
\begin{figure}[H]
\includegraphics[width=18cm]{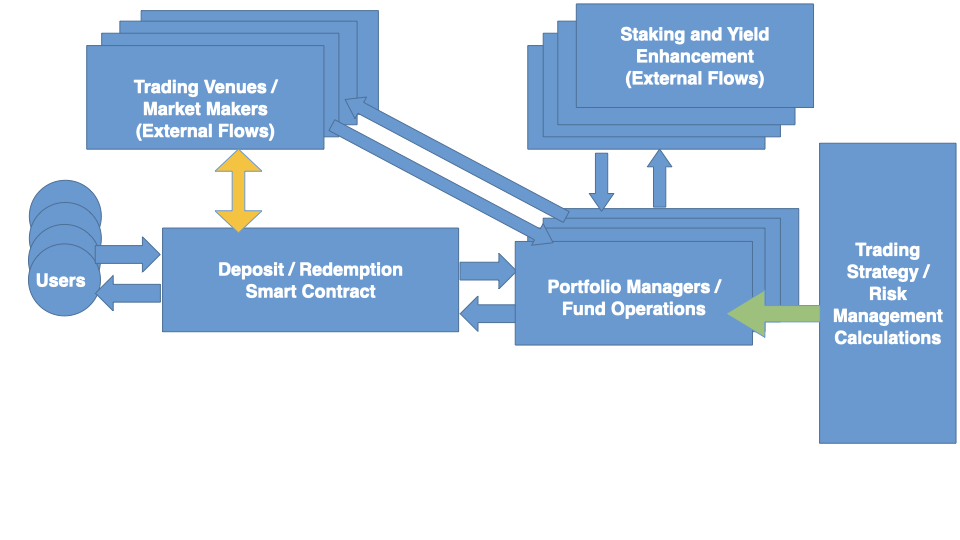}

\caption{\label{fig:Fund-Flow-Diagram}Blockchain Wealth Management: Fund Flow
Diagram}
\end{figure}

\end{doublespace}

Figure (\ref{fig:Fund-Flow-Diagram}) illustrates some common scenarios
regarding how fund flows need to happen for wealth management in a
decentralized environment. The blue arrows indicate fund flows in
Figure (\ref{fig:Fund-Flow-Diagram}). We can understand the diagram
in Figure (\ref{fig:Fund-Flow-Diagram}) by looking at the following
components - and steps - of the architecture. 
\begin{enumerate}
\item \label{enu:Investors-Deposit}Investors - or Users or Community Members
- deposit stable tokens via a Graphical User Interface - likely to
be web based - to purchase tokens in the investment fund (Ante, Fiedler
\& Strehle 2021; Lyons \& Viswanath-Natraj 2023; Jansen 1998; Oulasvirta
et al., 2020; End-notes \ref{enu:Stablecoin}; \ref{enu:The-graphical-user}). 
\item \label{enu:Investors-Exit}Similar to Point (\ref{enu:Investors-Deposit})
investors can exit the fund by sending in their fund tokens and getting
stable tokens back in exchange. A detailed description of how cryptocurrencies
other than stable tokens can be deposited - or redeemed - to participate
in - or exit - any investment fund is given in Kasaliya (2022).
\item \label{enu:Investors-authorize-transactions}Investors authorize transactions
using a blockchain wallet that manages their keys - and blockchain
addresses - to send and receive tokens (He et al., 2018; Karantias
2020; Suratkar et al., 2020; Houy et al., 2023; End-note \ref{enu:A-cryptocurrency-wallet}).
Deposits and Redemptions happen by interacting with a smart contract
- part of the fund infrastructure - that performs the calculations
and authentication related to ensuring that tokens transfers happen
according to fund prices and investor addresses.
\item The price of the fund - or the exchange rate that determines how many
fund tokens will be received in return for the stable tokens deposited
- will depend on the holdings within the fund and the current market
price of the assets held in the fund. The fund price will change as
the portfolio positions are rebalanced to meet risk and return objectives
and as the prices of the assets in the portfolio change. The nuances
of rebalancing crypto portfolios while responding to risk mitigation
needs are discussed in Kareem (2021-I); Kareem (2021-II); Kasaliya
(2022).
\item \label{enu:Portfolio-managers--}Portfolio managers - and others personnel
responsible for operating the fund - use the collected stable coin
proceeds to purchase other crypto-assets using Automated Markets Makers
or centralized exchanges or other trading venues (End-notes \ref{enu:AMM-Liquidity-Pairs};
\ref{enu:Centralized-cryptocurrency-excha}; \ref{enu:Decentralized-exchanges-(DEX)}). 

Fund personnel will need to have access to the smart contract keys
to be able to withdraw funds to buy or sell assets. It would be extremely
difficult to put the functionality to interface with external trading
venues into the smart contract that collects funds from investors.
This tenuous possibility is indicated by the yellow arrow in Figure
(\ref{fig:Fund-Flow-Diagram}). This issue is due to the difficulty
regarding deploying new versions of smart contracts and making changes
to blockchain infrastructure (Zou et al., 2019; Chen et al., 2021).
If a new trading venue becomes available up - or if any changes are
made to the application programming interface (API) of an existing
venue - it will require deploying a new version of the smart contracts
(Meng et al., 2018; Ofoeda et al., 2019; End-note \ref{enu:An-application-programming-interface}). 

It is possible to have an additional smart contract that keeps changing
- as the interface to external venues change or a component that changes
as the world changes while the rest of the infrastructure can remain
relatively unaffected - and connect this external interface smart
contract to the smart contract that manages investor deposits, redemptions
and holds funds. But even in this case - even if the issue of external
connections is resolved with a changing interface - trading personnel
will need to be able to access the funds to make the necessary fund
movements. 
\item \label{enu:Yield-Enhancement}The assets held in the portfolio can
be utilized towards yield enhancement services to generate additional
returns for the fund investors (Xu \& Feng 2022; End-note \ref{enu:Types-Yield-Enhancement-Services}).
Similar to Point (\ref{enu:Portfolio-managers--}), fund personnel
need to have the keys to be able access portfolio assets to delegate
them to external yield providers. 
\item Even though all the transactions happen on the blockchain, a large
off-chain software infrastructure - that reads information from the
blockchain transaction record - can be useful to provide inputs to
portfolio managers - and risk personnel - in terms of trading strategy
signals and risk metrics. This receipt of information is indicated
by the green arrow in Figure (\ref{fig:Fund-Flow-Diagram}).
\end{enumerate}

\subsection{\label{subsec:Fund-Loss-Categories}Fund Loss Categories}

\textbf{\textit{The main - and only - threat for any DeFi protocol
is simply the loss of funds. The complications are of course the many
ways in which this loss can happen. }}We initially broadly outline
the fundamental reasons for why - and how - funds can be lost. We
then zoom into our specific system components and consider the ways
an attack can happen.

We can categorize deficiencies in software wealth management - not
limited to blockchain alone - that can lead to loss of funds into
three categories: 1) Economic Principles; 2) Software Malfunctions;
3) Unlicensed Access. 
\begin{enumerate}
\item \label{enu:Economic}Loss of funds due to limitations in the economic,
financial and risk mitigation principles - used to build the corresponding
software system - can occur in several ways. 
\begin{enumerate}
\item A classic example of this type of catastrophe is the LUNA / UST failure
on the Terra protocol (Lee et al., 2022; Uhlig 2022; Briola et al.,
2023; End-note \ref{enu:Terra-May-2022}). The blowup of this network
happened due to failure of the arbitrage arguments in the stable coin
algorithm - which assume that participants will step in to correct
the price of the stable coin using the protocol token when there would
be a deviation in the peg of the stable coin (Kereiakes et al., 2019;
Brown \& Werner 1995; Shleifer, A., \& Vishny 1997; End-note \ref{Arbitrage, Wikipedia Link}). 
\item While it is easy to point out such mistakes after they have occurred
- perfect vision in hindsight - these anomalies are hard to detect.
Even if the designers of the system might be aware of these limitations,
such issues are especially obscured from the view of the users - or
investors - of the system. The failure of long term capital management
is another example when liquidity of the financial instruments - held
in the portfolio - dried up leading to a collapse of the fund (Lowenstein
2001; Edwards 1999; Jorion 2000; MacKenzie 2003; End-note \ref{enu:Long-Term-Capital-Management}).
\item Methods to bolster risk mitigation approaches are outlined in Kasaliya
(2022). While the intuition behind the economic concepts can seem
straightforward, these issues keep repeating themselves time and again
due to several fundamental advancements that are needed in understanding
the nature of uncertainty and human decision making (Reinhart \& Rogoff
2009; Karbeer 2016). Kashyap (2024) has specific guidelines for risk
mitigation in a cryptocurrency setting based on several situations
- and corresponding lessons - from traditional wealth management.
\end{enumerate}
\item \label{enu:Software-and-technical}Software and technical glitches
can be a lot more complicated to uncover due to the concealed nature
of several system components. The principles of thorough software
development, testing and maintenance have to be utilised. For the
sake of brevity, we have focused on the central elements of our technique
in Section (\ref{sec:The-Safe-House-Combination:}). The actual technical
implementation will have to cover several specialized scenarios, nuances
or other constraints. Additional checks pertaining to division by
zero and other such cases need to be considered in the software (End-note
\ref{enu:Software-Testing-Validation}). 
\begin{enumerate}
\item The tools we have outlined in Section (\ref{sec:Introduction:-DeFi-Defense})
should help with identifying blockchain specific code deficiencies.
The creation of better software development environments - and utilities
- for the blockchain space comparable to other software areas should
help immensely - including customized IDEs for blockchain software
development (Marchesi et al., 2018; Bosu et al., 2019; End-note \ref{enu:An-integrated-development-environment}). 

Chakraborty et al., (2018) find that standard software engineering
methods including testing and security best practices need to be adapted
with more seriousness to address unique characteristics of blockchain
and mitigate potential threats. Vacca et al., (2021) present solutions
to tackle software engineering-specific challenges related to the
development, test, and security assessment of blockchain-oriented
software. Qiu et al., (2019) propose a novel cloud-based solution,
namely ChainIDE, for the development of blockchain-based smart contracts
on multiple kinds of blockchain systems. 
\item Interestingly - but not surprisingly - blockchain systems can be used
to improve the software development life-cycle. Yilmaz et al., (2019)
consider most software failures as Byzantine failures and propose
a test-driven incentive mechanism - based on a blockchain concept
- to orchestrate the software development process where production
is controlled based on the working principles of blockchain. Farooq
et al., (2022) address ways to overcome the major issues of transparency,
trust, security, traceability, coordination, and communication in
Distributed Agile Software Development (DASD) by utilizing blockchain
technology (End-note \ref{enu:Distributed-agile-software}).
\item Human errors - both, while creating and using the software system
- can be considered as a subcategory within this group (Norman 1983;
Huang et al., 2012; Huang \& Bin 2017). 
\item Logical errors in the software code can also be viewed as falling
under this category (Nakashima et al., 1999; Engler et al., 2001;
Holzmann 2002). Clearly human errors in designing the system can contribute
to bugs in the code.
\item Losses when transferring funds via blockchain bridges also can be
categorized as software related (Belchior et al., 2021; Hafid et al.,
2020; Zhou et al., 2020; Stone 2021; Li et al., 2022; End-note \ref{enu:Blockchain-Bridges}).
Here we are restricting bridge transfer failures to be errors in the
functioning of the software that connects different blockchain networks.
Many bridge transfer issues happen due to authentication related issues
- or failures related to key management - pertaining to the operators
of the bridge. The unlicensed access issues we discuss next are limited
to the operation of the investment fund alone, without considering
unlicensed access to funds - being transferred on a bridge - that
are dependent on the keys for operating the bridge getting compromised.
Kashyap (2023) gives some suggestions on how to minimize fund transfer
needs in the current environment of severe bridge bottlenecks.
\end{enumerate}
\item \label{enu:Unatuhorized-Access}Zooming into DeFi wealth management,
all the funds - assets and tokens in the portfolio - will be maintained
on a decentralized system - either a smart contract or wallet - at
all times. This provides cryptographic assurance - based on public
/ private key principles - that no one can take the funds unless they
have the keys to access the funds. 

Hence, - in addition to the two categories of losses that we have
outlined above and even if due care is taken to ensure that proper
economic principles and software procedures are used - the most troubling
type of losses can occur due to unlicensed withdrawal of funds. This
presents a huge risk to blockchain funds - and the growth of decentralized
wealth management - wherein the entirely of the funds being managed
can be lost. 

There are three main types of threats a blockchain investment fund
will face in this scenario:
\begin{enumerate}
\item \label{enu:Withdrawal-of-funds-Unauthorized}Withdrawal of funds by
someone not authorized to access the system. The several ways in which
fraudulent activity happens in traditional finance - or other digital
arenas - can happen with respect to blockchain systems. This essentially
boils down to trying and figuring out someone's password - or dupe
them to reveal it - for any information technology system (Chiew et
al., 2018; Alabdan 2020; Desolda et al., 2021; End-note \ref{enu:Phishing-is-a}). 

In the blockchain realm, wallet applications have a password and once
that is divulged, the private keys saved in the wallet are compromised
(Andryukhin 2019; Pillai et al., 2019; Chen et al., 2020). Barkadehi
et al., (2018) identify different types of authentication systems
and provide information in terms of how such systems work, their usability
and drawbacks. Halgamuge (2022) develops a probabilistic model to
estimate the success probability of a malicious attacker. Huang et
al., (2011) propose a service to use one-time passwords to prevent
password phishing attacks. Kirda \& Kruegel (2006); Purkait (2012)
discuss how to protect users against phishing attacks. Wu et al.,
(2020) propose an approach to detect phishing scams on Ethereum by
mining its transaction records (End-note \ref{enu:Ethereum-is-a}).

In a blockchain setting, fund personnel can make this excuse that
their computer - or other devices - got hacked and the unauthorized
withdrawals were a result of the corresponding incident. 
\item \label{enu:Withdrawal-of-funds-Size}Withdrawal of funds by someone
who is authorized to access the system, but the amount withdrawn is
larger than the amount authorized. As discussed in Section (\ref{subsec:Components-of-Blockchain}:
Points \ref{enu:Portfolio-managers--}; \ref{enu:Yield-Enhancement})
fund personnel will need to have access to smart contract keys to
be able to transfer funds - to trade or participate in yield enhancement
platforms. Numerous instances of abuse of authority - violation of
risk limits - in extremely fortified financial institutions - despite
the many rigorous security procedures that these organizations followed
- are documented in Greener (2006); Krawiec (2000; 2009; 2021); Ouriemmi
\& Gérard (2023). In particular, most funds have strict limits on
the size of trades, positions and the corresponding fund flows. There
are several well chronicled accounts of employees - turning rogue
- having traded several multiples of the dollar values they are authorized
to trade, resulting in significant losses and sometimes even causing
entire organizational blowups.
\item \label{enu:Withdrawal-of-funds-Time}Withdrawal of funds by someone
who is authorized to access the system, but the time of withdrawal
is not when withdrawals are authorized. This happens when personnel
- with access to smart contract keys - trade using fund money at other
times when their trades are not necessarily the result of official
portfolio management objectives - but perhaps to take advantage of
some fancy schemes, that have not been approved, the traders or portfolio
manager come up with. This is similar to the case discussed above
(Point \ref{enu:Withdrawal-of-funds-Size}) about withdrawals being
larger than permitted limits. Over a period of time, such trades can
end up being large positions leading to huge losses for the fund.
Several examples are given in the references listed under Point (\ref{enu:Withdrawal-of-funds-Size}).
\end{enumerate}
\end{enumerate}

\subsection{\label{sec:Centralized-Risk:-Watching}Centralization and Rogue Risk
Mitigation: Watching Whales and Wallet Withdrawals}

As discussed in Section (\ref{subsec:Fund-Loss-Categories}), the
central fund repository - wallet or smart contract - might get hacked
when someone external to the organization either gets access to the
keys or figures out a way to steal funds when funds are required to
be moved to external venues or yield enhancement platforms by fund
personnel. The other possibility is that someone internal to the organization
might have turned rogue. 

There are two types of legitimate withdrawals that can happen: 
\begin{enumerate}
\item \label{enu:The-first-Investor}The first is by someone who has invested
money into the investment fund and is withdrawing the tokens that
rightfully belongs to them. 

Investors can get tricked to revealing their keys as discussed in
Section (\ref{subsec:Fund-Loss-Categories}). Another possibility
is that investors can get duped into believing that fake tokens are
indeed the authentic tokens issued by the fund - and even entire distributed
applications and platforms are real - and hence lose their investment. 

Chen et al., (2019) develop a method to automatically detect inconsistent
behaviors resulting from tokens deployed in Ethereum by comparing
information from three different sources, including the manipulations
of core data structures, the actions indicated by standard interfaces,
and the behaviors suggested by standard events.Gao et al., (2020)
have identified 2, 117 counterfeit tokens that target 94 of the 100
most popular cryptocurrencies by analyzing over 190,000 tokens (or
cryptocurrencies) on Ethereum. Zheng et al., (2022) propose a fake-token
detection algorithm to identify tokens generated by fake users or
fake transactions and analyze their corresponding manipulation behaviors. 

Xia et al., (2020) characterize several cryptocurrency exchange scams
by identifying over 1,500 scam domains and over 300 fake apps. Fake
accounts creating fake transactions to create the aura of authenticity
is also a real possibility (Huang et al., 2020; Rabieinejad et al.,
2023) including exchanges that fake trading volume to gain greater
credibility (Chen et al., 2022). This is despite the possibility that
blockchain services can serve to provide reliable methods of detecting
and preventing fake documentation - and other digital content such
as news - since information is made public and is accessible by anyone
(Hammi et al., 2021; Qayyum et al., 2019; Sayed 2019; Fraga-Lamas
\& Fernandez-Carames 2020).
\item \label{enu:The-second-type-trader}The second type of authorized withdrawal
happens when assets are added or removed from the wallet - using the
amount invested by investors or to redeem investors - for trading
them or to to delegate them to external yield enhancement platforms. 

These withdrawals involve interacting with other smart contracts such
as: constant market maker liquidity pools (LPs: Xu et al., 2021; Mohan
2022; End-note \ref{enu:Types-Yield-Enhancement-Services}). Centralized
exchange APIs might also fall under this category, but it would be
helpful to distinguish them from LPs and further technical research
should present ways of doing this. 
\end{enumerate}
We would like to put in place mechanisms that can mitigate the risk
of having one central wallet that holds the invested funds. Any additional
security measures need to ensure that the above two categories of
legitimate transactions - Points (\ref{enu:The-first-Investor}; \ref{enu:The-second-type-trader})
- are not duly inconvenienced. Otherwise, it would be hard to ensure
that portfolio management can happen seamlessly on blockchain. Hence,
we wish to prevent unauthorized withdrawals without introducing additional
constraints - and causing inefficiencies - for participants. We discuss
four broad mechanisms below to handle the risk of unauthorized withdrawals.
\begin{enumerate}
\item Quite simply we could have multiple wallets storing the portfolio
funds. Spreading the funds - and hence the keys - would make it slightly
harder for hackers to empty out all the fund storage pools. But, if
fund personnel have all the keys - corresponding to different storage
wallets or smart contracts - bundled together in the same wallet on
the same device, it would take the same amount of efforts for hackers
to steal funds in this case. Hence, the drawback in this scenario
is that fund personnel need to have different devices - to hold different
keys - to manage different storage mediums. 

Another related possibility here is for wallet management software
to require different password for keys corresponding to different
addresses - which would then reduce the need to have multiple devices.
But this would still require fund personnel to juggle multiple accounts
and passwords. Also, if one device gets compromised it is possible
that all the passwords on that device can be leaked and hence it is
safer to use multiple computing devices to access multiple accounts.
\item Having a Multiple-Signature wallet - (multi-sig: Bellare \& Neven
2007; Aggarwal \& Kumar 2021; Han et al., 2021; Zhang et al., 2022;
Goel et al., 2023; End-note \ref{enu:Multi-Sign Wallet}), which requires
every withdrawal transaction from the wallet to be signed by multiple
parties - will remedy unauthorized withdrawal risk to a great extent.
The tradeoff is that it is expensive in terms of the gas fees to have
every transaction being signed by many signatories. 

This step also slows down the overall system - by creating dependencies
on multiple parties - since buying and selling of assets will not
happen quickly. Pending fund transfers could get stuck for long time
periods when the required signatures are not obtained in a timely
manner. This can be extremely costly from a risk management point
of view, especially when assets need to be sold depending on negative
sentiment that can cause drastic - and rapid - drops in prices. A
similar situation affects assets buys and though this does not create
losses for the portfolio, it will affect fund returns when timely
buy trades are not made.

To remedy the timeliness aspect, we can necessitate that only withdrawals
above a certain amount require multi-sign functionality before they
are authorized. This is discussed further in Point (\ref{enu:X-Y-Withdrawal}).
But clearly this additional check introduces limitations that can
be exploited by making multiple smaller withdrawals. This approach
would also be highly restrictive in terms of having multiple parties
required to periodically check the withdraw queue to perform the necessary
approvals.
\item A white list - or an approved safe address list - such that withdrawals
can only be made to the addresses on this list can be helpful (Banerjee
et al., 2018; Laurent et al., 2018; Alam et al., 2022; End-note \ref{enu:A-whitelist-is}).
Usage of white lists is a common practice in many blockchain systems
especially when it comes to issuing new tokens or coins (Feng et al.,
2019). The white list scheme is also used in many software systems
to detect - and neutralize - phishing attacks (Jain \& Gupta 2016;
Varshney et al., 2016; Zaimi et al., 2020; Azeez et al., 2021).

Clearly the situation we are concerned with are write transactions
- which happen when funds are withdrawn - that change the state of
the decentralized ledger as opposed to read transactions which simply
check the state of the ledger (Weber et al., 2017; Baliga et al.,
2018; Abdella et al., 2021; Thakkar et al., 2018). All smart contracts
and addresses - that the fund wallet interacts with - need to be pre-approved
and present on the white list. Essentially a smart contract can also
be viewed as an address on blockchain. So when a new set of addresses
have to be added to the whitelist, multiple signatures are needed
to approve their inclusion into the whitelist.

Investor address inclusion in the white list can also be made automatically
- without the multi-sign feature - by adding any addresses that deposits
money to the fund wallet to the whitelist programmatically. In this
case, a rogue could deposit a small amount to get his - her / its
- address added to the white list and withdraw a large amount. We
articulate one method to handle this issue in Point (\ref{enu:X-Y-Withdrawal}). 

In addition a time lock can be added on the white list such that changes
to the list cannot be made for a certain number of hours after any
new addresses are added or removed (Lai et al., 2021; Bacis et al.,
2021; Mohanty \& Tripathy 2022; End-note \ref{enu:A-timelock-is}).
And an additional condition - utilizing time locks - can be that once
an address is included, withdrawals cannot be made to that new address
until a certain amount of time has elapsed. This is to ensure that
hackers cannot get onto the white-list and immediately withdraw funds.
If sometime has to be elapse - between inclusion of an address to
the whitelist and subsequent withdrawal to that address - and there
is more opportunity that the new addresses will get noticed and appropriate
action will be taken.
\item \label{enu:X-Y-Withdrawal}We can set rules in the smart contract
such that withdrawals from any one address above a certain amount,
$X$ USD, within a certain time, $Y$ minutes, will be flagged for
greater scrutiny. This additional step - when a transaction is flagged
- might be the use of multiple signatures to allow the withdrawal
from proceeding further. A clever hacker might withdraw to several
addresses and still take a large sum of money within a short span
of time. 

Hence we need to set a pre condition that the net deposits from an
address have to be above a certain amount, $Z$ USD, otherwise the
withdrawal to that address will be flagged for additional scrutiny.
Once the total deposits from an address are above $Z$ USD, then the
next condition - withdrawal above a certain amount, $X$ USD, within
a certain time, $Y$ minutes - will kick in and select transactions
for closer inspection. 

Having these rules are also helpful to ensure that the protocol will
not suddenly have a bank run type of situation (Uhlig 2010; Calvo
2012; Brown et al., 2017; Dijk 2017; End-note \ref{enu:A-bank-run}).
These rules are also helpful to ensure that huge withdraw requests
- being watchful of investors holding large positions, also known
as whales (Herremans \& Low 2022; Manahov 2022; End-note \ref{enu:A-cryptocurrency-whale,})
- do not lead to the need to liquidate large portions of the fund
rapidly. Such large withdraw requests can cause both existing investors
and the exiting investors to suffer losses due to the fall in prices
that happen when quick sales are made. Rules for an orderly mechanism
to handle outflow requests - and to ensure prices do not collapse
abruptly - are detailed further in Kareem (2021-II). Handling both
the inflow and outflow of funds with some type of queuing - so that
the need to make a lot of trades in a short amount of time is reduced
- is beneficial since it reduces implementation shortfall or slippage
(Perold 1988; Bertsimas \& Lo 1998; Karastilo 2020; End-note \ref{enu:Implementation-Shortfall}).

This mechanism can also be provided additional protection by combining
it with the white list feature. This would mean that any changes -
especially addition of new addresses - to the withdraw white list
will require multiple signatures before it can be approved. If an
address is not in the white list, no withdraw will happen. Only if
an address is in the white list, the additional checks mentioned in
this point will kick in providing greater security.
\end{enumerate}
Clearly, several of the above features can be combined in different
ways depending on the specifics of the system we are interested in
creating. A mechanism that combines features of both multiple signatures
and using other rules - including a type of password protection -
is described next in Sections (\ref{sec:The-Safe-House-Combination:};
\ref{subsec:The-One-Time-Next-Time-Password-}).

\section{\label{sec:The-Safe-House-Combination:}The Safe-House Safety Combination}

\subsection{\label{subsec:The-Safe-House-Architecture}The Safe-House Architecture}

\begin{doublespace}
\noindent 
\begin{figure}[H]
\includegraphics[width=18cm]{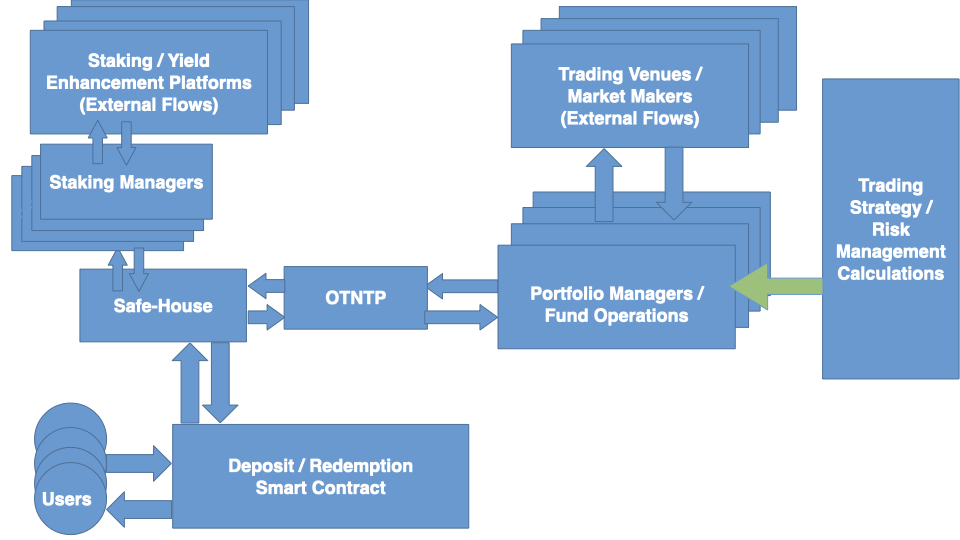}

\caption{\label{fig:Safe-House Architecture}Blockchain Wealth Management:
Safe-House Architecture}
\end{figure}

\end{doublespace}

The additional components in Figure (\ref{fig:Safe-House Architecture})
- compared to Figure (\ref{fig:Fund-Flow-Diagram}) which has the
basic components for blockchain wealth management - are briefly described
below. Later sections have detailed explanations regarding how each
of these additional components function.
\begin{enumerate}
\item The safe-house is an additional smart contract that interacts with
the smart contract that interfaces with investors to accepts deposits
and perform redemptions. The Safe-House enforces additional rules
when funds are transferred and this is discussed in detail below. 
\item The One-Time-Next-Time-Password (OTNTP) portion - explained in Section
(\ref{subsec:The-One-Time-Next-Time-Password-}) - sits between the
Safe-House and the Portfolio Mangers and performs identity verification.
\item The Staking Manager - which connects to the Safe-House is elaborated
in Section (\ref{subsec:Yield-Enhancement-Through}) - will be used
to delegate assets to external yield enhancement platforms.
\end{enumerate}

\subsection{\label{subsec:Multiple-Signatures,-One-Time-Ne}Multiple Signatures,
Encrypted Password Withdrawal and Counter Deposit Criteria}

In this section - and later sub-sections - we describe a methodology
that brings greater security to the protocol by combining: multiple
signature functionality for governance changes, encrypted password
authentications for each withdrawal and a counter deposit of equivalent
value to be made after each withdrawal - or after withdrawals reach
a maximum limit. 
\begin{lyxalgorithm}
The following points - and corresponding steps with explanations -
capture the methodology that will be required to significantly enhance
the overall security of the overall investment fund using the additional
components in Figure (\ref{fig:Safe-House Architecture}).
\end{lyxalgorithm}
\begin{enumerate}
\item The smooth functioning of any strategy will rely on two main roles:
Owners and Managers. 
\begin{enumerate}
\item Owners are responsible for all aspects of the fund and hence they
have the main fiduciary responsibility in terms of accepting money
from investors. The owners will own all the smart contracts and also
make changes to the key governing parameters. The owners will also
make the necessary changes to the governing parameters using multi-sig
wallets. An example is given in Point (\ref{Any-changes-to}). Clearly,
the greater the number of owners, the more secure the fund. But the
greater the number of owners - and hence the number of signatures
needed to make critical changes - the greater the time delays or other
corresponding inefficiencies.
\item Managers will do the day to day operational duties of the fund. Managers
will ensure that the smart contracts are operating properly, there
are enough funds to redeem investors and also invest the collected
funds into various assets. The main tasks and responsibilities of
managers are elaborated in Kareem (2021-II). We could have different
managers for different strategies - or sub-funds - within the overall
fund. The architecture given in Figure (\ref{fig:Safe-House Architecture})
can be replicated across the different sub-units to obtain greater
security.
\end{enumerate}
\item The main smart contract, $MAINSC_{t}$, of each strategy will receive
funds from investors for investment and it will also redeem investors
when they exit the strategy. 
\begin{enumerate}
\item The smart contract, $MAINSC_{t}$, will periodically move all funds
received to a safe house, $SAFEHOUSE_{t}$, which could be a wallet
address or another smart contract. 
\item The smart contract, $MAINSC_{t}$, will also take funds from the safe
house - or will be sent funds by a manager through the Safe-House
or otherwise - $SAFEHOUSE_{t}$, to redeem investors as necessary. 
\item The subscript $t$ denotes the state or address of the corresponding
component at time $t$ when there is no ambiguity. 
\item The state of the $SAFEHOUSE_{t}$ at any time will be either $OPEN$
or $CLOSED$.
\end{enumerate}
\item The manager, $MANAGER_{it}$, will access funds - stable coins or
other tokens - from the $SAFEHOUSE_{t}$, using a password unique
to the manager $i$ at time $t$. 
\begin{enumerate}
\item The password, $PASSWORD_{it}$, will be stored in - or accessible
to - the safe house, $SAFEHOUSE_{t}$, in an encrypted manner. 
\item To make a withdraw of funds or assets, $MANAGER_{it}$, will enter
the password, $PASSWORD_{it}$, using an internal only application
component or user interface. 
\item When the Safe-House is able to match the entered password to its encrypted
counterpart - stored in the Safe-House - only then the Safe-House
will open and allow funds to be taken by the manager. The manager
will then invest the funds in various assets, which are then sent
back directly to the $SAFEHOUSE_{t}$.
\item A certain nuance that arises here - specific to blockchain and decentralization
- is that the password sent to verify the identify of the manager
will be recorded as part of the transaction input and stored on the
ledger. This is necessary since the input sent to any smart contract
function to change the state of the system needs to be stored so that
the state change effected by that transaction can be verified by anyone.
This being the core requirement of blockchain - so that there is transparency
- gives way to the necessity to store the data that effects state
changes and make it publicly available. Hence, the password sent by
the manager has to be different each time since it will become publicly
known after a transaction is validated and the corresponding information
becomes public. A simple workaround can achieve the desired result
using a scheme we call the One-Time-Next-Time-Password, which is discussed
in Section (\ref{subsec:The-One-Time-Next-Time-Password-}).
\item Using the encrypted password protection scheme, it can be verified
with certainty who made the withdraw since the password used to make
the withdraw will be recorded. Hence if any manager turns rogue, they
can be identified and suitable measures can be initiated. To ensure
greater security and to prevent the additional costs, hassles and
publicity that might come with pursuing a rogue scandal, we can limit
the amount of funds or assets that can be taken in any one particular
transaction. This also ensures that the maximum damage that can be
done by rogue actions is minimized. In this case, we set a maximum
amount that can be taken out in one particular withdraw transaction,
$MAXOUTSAFEHOUSE_{t}$.
\end{enumerate}
\item \label{enu:In-addition,-withdraw-}In addition, after a withdraw is
made for any particular amount - less than the maximum permissible
amount $MAXOUTSAFEHOUSE_{t}$ - a counter deposit has to be made that
is comparable in value to the amount withdrawn, $WITHDRAW_{it}$,
by the manager $i$ at time $t$. 
\begin{enumerate}
\item This counter deposit has to be made from time $t$ - when the withdraw
is made - until a certain time period after time, $t$ - that is $t+CDTIME_{t}$. 
\item Here $CDTIME_{t}$ denotes a time period measured in seconds. The
cumulative value of the counter deposits made during the interval
from $t$ to $t+CDTIME_{t}$ is denoted by $COUNTERSAFEHOUSE_{t,t+CDTIME_{t}}$.
Clearly, for simplicity, $CDTIME_{t}$ can be a running time counter
starting from time $t$ until a later time. The absence of one clock
- or a standard time counter - for everyone is a limitation that is
still being overcome in the decentralized realm. Even though, getting
the same time for everyone is still hard to accomplish for all participants,
several alternatives are being pursued. One approach is to use the
time to validate blocks or the block-time as a reasonably proxy (Swan
2016; Zhang et al., 2019; Ladleif \& Weske 2020; End-note \ref{enu:The-block-time}).
\item $DEPOSIT_{jt}$ represents the deposit made at time $t$ by address
$j$. If the difference between the withdrawal made at time $t$,
$WITHDRAW_{it}$, and the cumulative value of the counter deposits
till a time after $t$, $COUNTERSAFEHOUSE_{t,t+CDTIME_{t}}$, - is
less than a certain threshold, $SAFEHOUSETOLERANCE_{t}$, the safe
house will remain closed for all further withdraws. The $SAFEHOUSETOLERANCE_{t}$
is best represented as a percentage figure and converted to a notional
value by multiplication with the withdraw amounts. Once the Safe-House
is closed after a withdraw and the necessary amount of counter deposits
have not been received - to open it again - intervention by the owners
of the Safe-House using multiple signatures will be required. 
\item It needs to be ensured that the withdraws made by the smart contract,
$MAINSC_{t}$, to redeem investors are not counted. Hence the withdraws
that matter are the ones from addresses other than the $MAINSC_{t}$.
Clearly withdraws from addresses that hold the manager - or owner
- role are the only ones that will be allowed to access the $SAFEHOUSE_{t}$.
It also needs to be checked that the counter deposits are not funds
received from investors through the smart contract, $MAINSC_{t}$.
Hence the counter deposits have to be received from addresses other
than the $MAINSC_{t}$. This feature can also be implemented by having
different functions in the Safe-House to interact with investors and
managers.
\end{enumerate}
\item \label{enu:Criterion-One}All of the above conditions are captured
- using the notation developed thus far in Point (\ref{enu:In-addition,-withdraw-})
- more formally below:
\begin{criterion}
\label{cri:Safe-House-Criterion-One-Inequalities}The following inequalities
and equations encapsulate the essence of the Safe-House,
\begin{equation}
\left|WITHDRAW_{it}\right|\leq MAXOUTSAFEHOUSE_{t}
\end{equation}
The Safe-House state will be closed, $SAFEHOUSE_{t}\equiv CLOSED$,
after a withdraw at time $t$ as long the following condition holds,
\begin{align}
\left|WITHDRAW_{it}\right|-\left|COUNTERSAFEHOUSE_{t,t+CDTIME_{t}}\right| & <\nonumber \\
\left|WITHDRAW_{it}\right|\left(SAFEHOUSETOLERANCE_{t}\right)
\end{align}
\begin{equation}
COUNTERSAFEHOUSE_{t,t+CDTIME_{t}}=\sum_{l=t,i,j\neq MAINSC_{t}}^{t+CDTIME_{t}}\left(DEPOSIT_{jl}\right)
\end{equation}
Note that with a slight abuse of notation, we do not include a separate
summation across all the managers and owners represented by the indices
$i,j$.
\end{criterion}
\item An equivalent way to accomplish the Criterion (\ref{cri:Safe-House-Criterion-One-Inequalities})
in Point (\ref{enu:Criterion-One}) is given below:
\begin{criterion}
A simpler alternative to the above - Criterion (\ref{enu:Criterion-One})
- would be to view withdraws as being allowed up to a certain maximum
limit. 
\begin{itemize}
\item When a withdraw a made, we deduct from the maximum limit and when
a deposit is made we add to the maximum limit. The maximum limit will
be the amount, $MAXOUTSAFEHOUSE_{t}$. 
\item The sum of the deposits can be less than the withdraws by a certain
tolerance amount. The tolerance amount is obtained by using the $SAFEHOUSETOLERANCE_{t}$
- which is best represented as a percentage figure - and converting
the percentage figure to a notional value by multiplication with the
withdraw amounts. 
\item We define a starting time to be when the safe house is opened for
the first time after being closed.Hence the difference of the cumulative
withdraws and deposits from any starting time, $T$ to any future
time, $T+\triangle T$ has to be lesser than the maximum limit adjusted
by the tolerance amount. Clearly we can start the counters from the
time the safe house is established and opened for business, so that
$T=0.$
\item A starting time is when the safe house is opened for the first time
after being closed. The Safe-House state will stay open, $SAFEHOUSE_{t}\equiv OPEN$,
after a starting time, $T$ until the time, $T+\triangle T$, as long
the following condition holds,
\begin{align}
\left\{ \sum_{t=T,i,j\neq MAINSC_{t}}^{T+\triangle T}\left[\left|WITHDRAW_{it}\right|-\min\left(MAXOUTSAFEHOUSE_{t},\left|DEPOSIT_{jt}\right|\right)\right]\right\} \nonumber \\
<MAXOUTSAFEHOUSE_{t}+\sum_{t=T,i,j\neq MAINSC_{t}}^{T+\triangle T}\left|WITHDRAW_{it}\right|\left(SAFEHOUSETOLERANCE_{t}\right)
\end{align}
\item Note that with a slightly abuse of notation, we do not include a separate
summation across all the managers and owners represented by the indices
$i,j$.
\end{itemize}
\end{criterion}
\item \label{Any-changes-to}Any changes to the governance parameters -
or strategic parameters - will require authentication with multiple
signatures from multiple parties - or owners. Examples for this are:
\begin{condition}
To set someone with a manager role. Once someone has been given a
manager role, they need to set a password for themselves in the safe-house
and confirm it again. 
\begin{condition}
To revoke someone from a manager role, any of the multiple signature
owners can do it without approval from others. This is to ensure timely
intervention when necessary.
\begin{condition}
To change parameters such as the $SAFEHOUSETOLERANCE_{t}$, $CDTIME_{t}$
and $MAXOUTSAFEHOUSE_{t}$. We outline some guidelines for setting
these variables.
\end{condition}
\begin{itemize}
\item The $SAFEHOUSETOLERANCE_{t}$ will be related to the slippage tolerance
discussed in Kareem (2021-II). 
\item An insurance - or risk reserve - will be maintained as part of the
fund treasury to finance various risk limits. The $MAXOUTSAFEHOUSE_{t}$
will need to be set as part of this risk reserve. 
\item The $CDTIME_{t}$ will depend on how long the off chain risk and portfolio
management infrastructure - scripts and other calculations with trader
/ portfolio manager oversight - take to complete. 
\end{itemize}
\begin{condition}
There are many other operational parameters that the manager might
need to be able to update at times. These are related to various fees
the fund needs to collects when deposits and redemptions happen (Kareem
2021-II). It is very important to be able to distinguish between governance
and operational parameters and decide who - managers or owners - has
responsibility to change those. If necessary, a detailed discussion
of operational tasks - and related parameters - is available separately
in Kareem (2021-II).
\end{condition}
\end{condition}
\end{condition}
\item The critical aspect with the counter deposit criteria is the verification
of the value of the counter deposits, $COUNTERSAFEHOUSE_{t,t+CDTIME_{t}}$. 
\begin{enumerate}
\item The counter deposit has to be made using only assets that have been
included in a pre-approved list. This is to ensure that counter deposits
are made with assets that satisfy due diligence requirements - as
mandated by the fund strategies - and the Safe-House cannot be simply
stuffed with junk assets to satisfy the corresponding criteria. The
pre-approved list of assets can be the list of assets included in
the portfolio of the corresponding strategy. 
\item A very reliable price provider has to be used for the counter deposit
value calculations. Oracles are a solution for smart contracts to
obtain inputs from the outside world (Beniiche 2020; Pasdar et al.,
2021; Lys \& Potop-Butucaru 2022; Pierro et al., 2020; 2022; Caldarelli
2022; Pasdar et al., 2023). Price oracles on several platforms aggregate
prices from various sources and provide values that cannot be easily
manipulated (End-note \ref{enu:Blockchain-oracles-are}). That said,
oracle price related hacks are possible attack vectors in DeFi (Al-Breiki
et al., 2020; Caldarelli 2020; Oosthoek 2021; Caldarelli \& Ellul
2021) and hence this particular component needs additional attention
when the systems are running. 
\item A safety measure involves setting a reference price for each asset
in the Safe-House. When withdraws - or deposits - are being made using
external prices that are significantly different from this reference
price, movement of those assets can be frozen until someone can investigate
this matter further. The reference price should only be updated by
owners and this is another example of a governance parameter. The
list of approved assets and their latest reference price must be reviewed
regularly - daily or even multiple times a day - by portfolio managers,
risk personnel and even the owners of the fund. 
\begin{criterion}
Using this condition the deposit made to the Safe-House can be written
as 
\begin{equation}
DEPOSIT_{it}=\sum_{k=1}^{K}\left(ASSETPRICE_{ikt}\right)\left(ASSETQTY_{ikt}\right)
\end{equation}
Here $K$ is the number of assets approved for manager - or owner
- , $i$. $ASSETPRICE_{ikt}$and \textup{$ASSETQTY_{ikt}$ represent
the asset price and asset quantity being deposited into the Safe-House
at time $t$. }Note that with a slight abuse of notation, we do not
include a separate summation across all the managers and owners represented
by the indices $i,j$. This summation of deposits across all owners
and managers gives the total counter deposit value, $COUNTERSAFEHOUSE_{t,t+CDTIME_{t}}$. 
\begin{criterion}
Withdraws can also be represented using a similar condition,
\begin{equation}
WITHDRAW_{it}=\sum_{k=1}^{K}\left(ASSETPRICE_{ikt}\right)\left(ASSETQTY_{ikt}\right)
\end{equation}
\end{criterion}
\end{criterion}
\item Portfolio assets will be staked in various farms, pools or yield generating
devices. The confirmation for staking in any farm or pool - which
are usually liquidity pool (LP) tokens - can be sent to the $SAFEHOUSE_{t}$.
The verification of value of these LP tokens can be a challenge, but
are being done currently by many tools - such as Nansen.AI (Moro-Visconti
et al., 2023; Zhu et al., 2023; End-note \ref{enu:Nansen-is-a}) -
across various networks. 

The value of the LP tokens can also be obtained from the yield enhancement
platform itself by invoking suitable functions- since essentially
the automated market maker can provide a reference price. But changes
to the API will involve quite a bit of maintenance for the fund technologists.
Also it is easy to manipulate the price of a single market maker. 

The discussion in Section (\ref{subsec:Yield-Enhancement-Through})
- using the Staking Manager - is an alternative to do yield enhancement
that bypasses the withdraw and counter-deposit criteria. This would
mean that LP tokens as confirmation for staking will be sent directly
to - and from - the Safe-House without having to go through the withdraw
- and deposit - aggregation.
\item The Safe-House contracts have been successfully deployed for commercial
use on multiple platforms - Ethereum, Binance and Polygon - and hundreds
of transactions have been performed without any issues. The use of
price oracles for counter deposit value verification - including the
use of a default reference price - has also been tested satisfactorily.
\end{enumerate}
\end{enumerate}

\subsection{\label{subsec:The-One-Time-Next-Time-Password-}OTNTP: The One-Time-Next-Time-Password}

The One-Time-Next-Time-Password (OTNTP) concept will be used to verify
the identity of the manager and to allow Safe-House access for making
withdrawals. It is important to note that the OTNPT is an additional
layer or protection and is not a substitute for the private keys being
used by fund personnel. This modified OTNPT scheme should help with
password protection in decentralized environments where all transaction
information has to be made public for verification purposes. 

The internal application or component that will be used to access
funds from the Safe-House, $SAFEHOUSE_{t}$, will be called the administrative
GUI (Graphical User Interface) or admin GUI for short. To access this
internal application - which will perform various investment duties
by interacting with the various smart contracts including the Safe-House,
$SAFEHOUSE_{t}$ - another administration password will be required.
We refer to this password for the admin GUI as the admin password. 

We denote the admin password for manager $i$ at time $t$, $MANAGER_{it}$,
as, $ADMINPASSWORD_{it}$. The One-Time-Next-Time-Password for manager
$i$ at time $t$ will be denoted as $OTNTP_{it}$. 
\begin{lyxalgorithm}
The main steps to implement The One-Time-Next-Time-Password (OTNTP)
are:
\end{lyxalgorithm}
\begin{enumerate}
\item To access the funds - or assets - in the Safe-House, $SAFEHOUSE_{t}$,
the $MANAGER_{it}$ using the admin GUI will need to send the raw
form or plain text version of the \textbf{CURRENT} One-Time-Next-Time-Password
corresponding to him, $OTNTP_{it}$ that is needed for the current
access. 
\begin{enumerate}
\item When sending the raw form of the current $OTNTP_{it}$, the admin
GUI of the $MANAGER_{it}$ will create the \textbf{NEXT} One-Time-Next-Time-Password,
$OTNTP_{it+\triangle t}$, encrypt it and send it to the Safe-House,
$SAFEHOUSE_{t}$. 
\item The encrypted form of $OTNTP_{it+\triangle t}$ will be saved in the
Safe-House so that it can be used to access the Safe-House, $SAFEHOUSE_{t}$
the next time by the $MANAGER_{it}$.
\end{enumerate}
\item The raw form of the next One-Time-Next-Time-Password, $OTNTP_{it+\triangle t}$
can be saved on the computing device of the manager, $MANAGER_{it}$
in a protected file, $PROTECTEDFILE_{it}$. 
\begin{enumerate}
\item The password to access the contents of the protected file, $PROTECTEDFILE_{it}$
is the $ADMINPASSWORD_{it}$ for manager, $MANAGER_{it}$. 
\item Clearly, the $OTNTP_{it+\triangle t}$ can also be stored on external
devices inside password protected files and access to this protected
file, $PROTECTEDFILE_{it}$ will require knowing the $ADMINPASSWORD_{it}$
for manager, $MANAGER_{it}$. 
\item Hence, even if the wallet of the $MANAGER_{it}$ gets hacked, without
knowing the $OTNTP_{it+\triangle t}$ or the $ADMINPASSWORD_{it}$
of the manager access to the $SAFEHOUSE_{t}$ for withdrawals is not
possible.
\end{enumerate}
\item The One-Time-Next-Time-Password, $OTNTP_{it}$, can simply be a random
number of sufficient length so that the number of tries for someone
trying to force access to the $SAFEHOUSE_{t}$ is probabilistically
very small. 
\begin{enumerate}
\item This also means that after a few number of wrong attempts using the
incorrect One-Time-Next-Time-Password, $OTNTP_{it}$, the $SAFEHOUSE_{t}$
will go into the $CLOSED$ state for everyone. 
\item Further intervention will be necessary from the owners using multi-sign
transactions to restore the state of $SAFEHOUSE_{t}$ for normal operation
and to be able to go back to the $OPEN$ state.
\item Having non-numeric characters is helpful to decrease the probability
of someone guessing the password.
\end{enumerate}
\item The scenario we have discussed above is for the regular operation
of the $SAFEHOUSE_{t}$ using the One-Time-Next-Time-Password, $OTNTP_{it}$
or for system when it has been operational for sometime. 
\begin{enumerate}
\item To start the operation of the system, the owners will need to seed
the encrypted form of the $OTNTP_{it}$ into the $SAFEHOUSE_{t}$. 
\item A simpler and more practical alternative would be for the manager
himself to seed the $SAFEHOUSE_{t}$ with the encrypted version of
the first One-Time-Next-Time-Password, $OTNTP_{it}$ and then use
this for the first withdrawal that follows. 
\item When the manager will seed the Safe-House for the first time, multiple
signature verification can also be enforced for increased security. 
\item The same seeding mechanism can be used to restore the state of $SAFEHOUSE_{t}$
for normal operation and to be able to go back to the $OPEN$ state,
after it gets $CLOSED$ due to multiple incorrect attempts.
\end{enumerate}
\item To obtain access to the $SAFEHOUSE_{t}$ would require gaining control
of the wallet of the manager and also hack his computing device -
and knowing his $ADMINPASSWORD_{it}$ - to obtain the contents of
the $PROTECTEDFILE_{it}$ - which is the OTNTP. 
\begin{enumerate}
\item This OTNTP feature combined with the counter deposit requirement ensures
that the effort to compromise the $SAFEHOUSE_{t}$ is significant
in comparison to the reward that can be obtained - which is the maximum
amount that can be taken from the Safe-House, $MAXOUTSAFEHOUSE_{t}$
without counter deposit verification. To steal assets above the maximum
Safe-House amount will require a lot of additional resources to manipulate
price oracles or other price feeds.
\end{enumerate}
\end{enumerate}

\subsection{\label{subsec:Yield-Enhancement-Through}The Staking Manager: Yield
Enhancement Through Third Party Delegation of Assets}

If we need to delegate assets from the Safe-House to third party yield
enhancement platforms, we will need a Staking Manager (SM) contract.
The main purpose of the SM is to handle the specifics of token management
with respect to the protocol - or DeFi platform - that is being used
for the yield enhancement. The reason for having a separate SM is
because third party platforms change their smart contract versions
through which they provide yield enhancement services. 

In addition, we will be seeking new yield enhancement platforms as
and when they arise. When such external interface changes are necessary,
we eliminate the need to modify and re-deploy the Safe-House. This
becomes possible since the SM contract can be modified to deal with
the changes and it can be deployed again and connected to the Safe-House.
Hence the SM will change as the external world changes but its connection
to the fund - that is to the Safe-House - will remain the same. By
linking different SMs that provide these external connections we can
retain the same Safe-House without any changes to it.

For now, we will consider the following three types of yield enhancement
services (Kiong 2021; Xu \& Feng 2022; Cousaert, Xu \& Matsui 2022;
End-note \ref{enu:Types-Yield-Enhancement-Services}): AMM Liquidity
Pairs (AMM LPs), LP Token Staking and Single Sided Staking. Other
types of yield generation protocols can be handled in a similar manner
(End-notes \ref{enu:DefiLlama-is-the}; \ref{enu:CoinMarketCap};
\ref{enu:Crypto-Ranking}). Examples of Single sided staking are CAKE
and BSW. Examples of AMM LPs are ETH-LINK, SOL-ETH on Uniswap and
ADA-BNB, DOT-BNB on Pancake swap (End-notes \ref{enu:Uniswap-is-a};
\ref{enu:PancakeSwap-(CAKE)-is}). Once we deposit the two tokens
corresponding to an AMM LP pair, we get the relevant LP token. This
LP token can then be staked in the corresponding staking pool to earn
the platform rewards such as Pancake swap tokens. 

The Safe-House needs to have the following functions which can be
invoked for any asset or pair of assets. The following functions can
also support more than two assets by designing the function interface
such that an array of assets can be passed to the function as an input.
The quantity - or number of tokens - also needs to be passed along.
Along with the list of assets and quantities, an instruction ID will
have to be passed to the function being invoked in the Safe-House.
The instruction ID can be an integer. The functions needed in the
Safe-House are:
\begin{enumerate}
\item Add Liquidity
\item Remove Liquidity
\item Stake
\item Un-Stake
\item Claim Rewards
\end{enumerate}
The portfolio manager will call one of the functions in the Safe-House
from the above list. The Safe-House will then delegate the call to
the Staking Manager - which will then do the needful and perform the
corresponding action by interacting with the third party contract
- depending on the instruction ID. The instruction ID will inform
the SM which external platform to interface and what specific activities
to carry out.
\begin{example}
Suppose we wish to add liquidity to an AMM LP then we will call the
add liquidity function - with the relevant parameters. The staking
manager then interacts with the third party service and the LP tokens
received by this add liquidity call will then be sent back to the
Safe-House. A similar mechanism happens when we need to remove liquidity
or for the other functions.
\end{example}
When we start staking a new pair - or a new token or when the third
party protocol changes its smart contract interface - the Staking
manager will change to be able to handle the corresponding modifications.
By restricting changes only to the Staking manager, the instruction
IDs and the set of actions corresponding to each instruction ID, the
Safe-House functions will not need to know about the nuances of staking
and hence they do not need to be changed. 

It is important to ensure that when tokens are received back - as
confirmation for depositing into external yield generation platforms
or when liquidity is removed or tokens are un-staked - they need to
be received back directly into the Safe-House. It is possible to received
the tokens back in the SM and then the managers can move the assets
between the Safe-House and Staking Manger as necessary. The return
address should be the Safe-House and only the owners using multiple
signatures can change this return address when the system is reinitialized
or any major changes are made to the components.

A related enhancement - which requires an additional transaction but
does not require redeploying the SM contract - will be to maintain
a list of staking manager contracts in the Safe-House. An interface
to the Safe-House will pick the right staking manager depending on
the particular asset being considered. So the assets can be transferred
to the corresponding staking manager contract, which will handle further
movements of funds to external platforms. The funds can be received
directly back into the Safe-House or through the SM as well similar
to the sending of assets. The mapping of assets to staking manager
contracts needs to be maintained and updated as staking protocols
change their entry point definitions.

\subsection{\label{subsec:Mitigation-of-the}Preventing the Three Types of Unauthorized
Withdrawals}

Here we quickly look at how the solutions in Sections (\ref{subsec:The-Safe-House-Architecture};
\ref{subsec:Multiple-Signatures,-One-Time-Ne}; \ref{subsec:The-One-Time-Next-Time-Password-};
\ref{subsec:Yield-Enhancement-Through}) can prevent the three types
of Unauthorized Withdrawals listed in Section (\ref{subsec:Fund-Loss-Categories};
Point \ref{enu:Unatuhorized-Access}).
\begin{enumerate}
\item \label{enu:Withdrawal-of-funds-someone}Withdrawal of funds by someone
not authorized to access the system. The combination of public-private
cryptographic keys - which is common to all blockchain transactions
- provides the first level of security support. The OTNTP protection
- the innovation we have added - provides an extra layer of identity
verification to prevent withdrawals even when private keys are compromised.
We next consider (Point \ref{enu:Withdrawal-of-funds-amount}) the
scenario when both private keys and OTNTP are obtained by a malicious
party.
\item \label{enu:Withdrawal-of-funds-amount}Withdrawal of funds by someone
who is authorized to access the system, but the amount withdrawn is
larger than the amount authorized. If an internal party - that is
the manager with access to the keys that can remove assets from the
Safe-House - turns rogue or an external actor has obtained the keys
- including the OTNTP - for making withdrawals, the maximum withdrawal
amount from the safe-house ensures that the maximum one-time loss
is limited. 

In blockchain terminology, if someone has the keys they are authorized
to do transactions. Hence by restricting the maximum withdraw amount
- and having a counter deposit criteria - we are able to ensure that
even if both the private key and OTNTP are hacked by someone, the
fund is protected from losing everything. It is important to remember
that Points (\ref{enu:Withdrawal-of-funds-someone}; \ref{enu:Withdrawal-of-funds-amount})
work in tandem to provide complete security for the investment fund.
\item \label{enu:Withdrawal-of-funds-Block-Time}Withdrawal of funds by
someone who is authorized to access the system, but the time of withdrawal
is not when withdrawals are authorized. This third type of unauthorized
access can be prevented by having withdrawals to depend on the passage
of certain lengths of block-times. That is we allow a window of withdrawals
based on the time to validate a certain number of blocks (End-note
\ref{enu:The-block-time}). For example, after a withdrawal a made,
we can prevent the next withdrawal from happening until a certain
number of blocks have been validated. Similarly we can specify that
once a withdraw has started all withdraws need to be completed - within
the maximum amount specified in Point (\ref{enu:Withdrawal-of-funds-amount})
- before a certain number of block-times have elapsed. 

This additional feature - using block-times - can be implemented to
add time delays that makes it cumbersome to steal money from the fund.
The more important point is that due to the delays - by having periodic
oversight: regular fund reporting and monitoring - wrongful acts and
actors can be detected. When fund personnel begin their work day -
or start a new shift when operating in shifts - they need to perform
a set of procedures to reconcile changes from the last time period
when legitimate transactions were made. Time delays - using the block-time
- ensures that there is sufficient time to perform the necessary reconciliation
checks and this can call attention to any unauthorized activity. This
step becomes an additional shield of safety that combines seamless
with Points (\ref{enu:Withdrawal-of-funds-someone}; \ref{enu:Withdrawal-of-funds-amount})
to ensure greater security. 
\end{enumerate}

\section{\label{sec:Areas-for-Further}Areas for Further Research}

A key policy variable is the amount of funds that can be withdrawn
from the Safe-House at any one point in time. This maximum amount,
that can be taken out in one particular withdraw transaction: $MAXOUTSAFEHOUSE_{t}$,
has strong implications for the operational efficiencies for the investment
fund. Clearly, this amount should be linked to any risk management
methodologies being used by the fund (Jorion \& Khoury 1996; Rasmussen
1997; Alexander 2005; Horcher 2011; Andersen et al., 2013). This maximum
one time loss has to be a risk management limit and should depend
on the overall assets under management (AUM: End-note \ref{enu:Finance-AUM})
or the total value locked (TVL: End-note \ref{enu:TVL}). 

If the maximum withdraw limit is low, it will increase operational
inefficiencies and further understanding this parameter - in terms
of the investment strategies and fund flow requirements - can be helpful.
A thorough analytical framework can be developed for setting this
maximum withdraw amount - and the corresponding fund transfer requirements
- in subsequent papers. Two analytical derivations can be very useful:
1) To estimate the blockchain transaction costs for different maximum
safe house amounts and by comparing them to the slippage costs when
trades of different sizes are made. When trade slippage costs are
more than blockchain transaction costs the maximum amount can be decreased
and vice versa. 2) The prices of assets will change between the time
of withdraw and subsequent deposits, hence the safe-house tolerance
needs to depend on this change in prices. Even if someone turned rogue
within the firm - and tried to benefit using the price movements between
withdraw and deposit transactions - the estimates on the amounts that
could be lost can be included in risk management reports as a precautionary
measure.

Having multiple Safe-Houses is also a possibility for different funds
and groups of assets. The pros and the cons are the maintenance requirements
of having to permission personnel for different sub-houses and also
ensuring that the right assets are allocated to different Safe-Houses.
Combining Safe-Houses ensures that when multiple funds have the same
assets, a certain amount of netting benefits will accrue. This can
reduce the number of transactions needed.

Hackers can try to override the OTNPT mechanism using a brute force
attack by trying to guess multiple times what the password might be.
Hence, it is prudent to put the Safe-House into a locked state when
a certain number of unsuccessful attempts are made. The number of
incorrect attempts - before the Safe-House gets locked - could depend
on the wealth under management, the maximum withdraw amount, transaction
costs - gas fees - on the platform and other risk management parameters.

A point of attack for the Safe-House is through price manipulation
of the asset prices in the Safe-House value verification mechanism.
These type of attacks are based on price changes triggered by fund
movements, related to taking large loans called flash loans, all happening
within the same transaction (Qin et al., 2021; Wu et al., 2021; Werapun
et al., 2022; Xue et al., 2022). Such attacks need additional enhancements
to the Safe-House, especially to monitor the price data sent to the
Safe-House verification mechanism from blockchain oracles (Adler et
al., 2018; Al-Breiki et al., 2020; Beniiche 2020; Caldarelli 2020;
Lo et al., 2020; Mammadzada et al., 2020; Caldarelli \& Ellul 2021;
Pasdar et al., 2022). 

Illiquid instruments which are not traded on many venues are particularly
sensitive to price manipulation attacks. Assets can be categorized
into different buckets based on market capitalization or trading volume
or the number of trading venues. Different withdraw limits can be
set for different asset categories. The problem could be compounded
for assets that do not even have price oracles available for them.
In such cases, getting prices directly from liquidity pools is possible.
But having only one source is a huge point of vulnerability. Another
approach would be to set a default price based on moving averages,
for the asset without price oracle feeds, and updated regularly as
the asset price changes.

\section{\label{sec:Conclusion}Conclusion}

Our innovation - which is created from first principles and readily
combined with existing protocols - is entirely custom built to safeguard
fund movement workflows in a blockchain environment and we call this
The Safe House. The Safe House is the combination of a novel software
engineering architecture combined with blockchain cryptographic security,
specific to handling fund movements, with certain multi-signatory
approvals required for changing key governance policies. This approach
will limit any potential one-time loss to a negligible amount. 

An additional enhancement called the OTNTP (one time next time password)
can add additional protection to the Safe-House reducing the one time
loss threshold as well. It is important to remember that the OTNPT
can be used only by someone authorized to act using wallet's - or
private key credentials - belonging to fund personnel. So after getting
access to an employee's keys the next obstacle involves trying to
figure out the OTNTP. And even if both safety mechanisms - OTNTP and
private keys - fail, the counter deposit criteria - including the
maximum withdraw amount - will ensure that the maximum loss is limited.
To overcome the counter deposit criteria will require expending lots
of resources to manipulate price oracles. 

Hence, our multi-layer defense mechanism makes it significantly harder
- than a regular DeFi project - for someone to successfully breach
through the different shields and get access to investment funds.
We have also discussed how transfer of assets can happen to external
yield generation protocols directly from the Safe-House in a secure
manner. The Safe-House and Staking manager contracts have been successfully
deployed for commercial use on multiple platforms - Ethereum, Binance
and Polygon - and hundred of transactions have been performed - as
part of daily operational usage - without any problems. The use of
price oracles for counter deposit value verification has also been
tested satisfactorily.

The combination of the Safe-House and the OTNTP will provide much
needed improvements to wealth managers that wish to operate entirely
on a decentralized environment. 

\section{\label{sec:References}References}
\begin{enumerate}
\item Abdella, J., Tari, Z., Anwar, A., Mahmood, A., \& Han, F. (2021).
An architecture and performance evaluation of blockchain-based peer-to-peer
energy trading. IEEE Transactions on Smart Grid, 12(4), 3364-3378.
\item Adler, J., Berryhill, R., Veneris, A., Poulos, Z., Veira, N., \& Kastania,
A. (2018, July). Astraea: A decentralized blockchain oracle. In 2018
IEEE international conference on internet of things (IThings) and
IEEE green computing and communications (GreenCom) and IEEE cyber,
physical and social computing (CPSCom) and IEEE smart data (SmartData)
(pp. 1145-1152). IEEE.
\item Aggarwal, S., \& Kumar, N. (2021). Attacks on blockchain. In Advances
in computers (Vol. 121, pp. 399-410). Elsevier.
\item Alabdan, R. (2020). Phishing attacks survey: Types, vectors, and technical
approaches. Future internet, 12(10), 168.
\item Alam, S., Zardari, S., \& Shamsi, J. A. (2022). Blockchain-Based Trust
and Reputation Management in SIoT. Electronics, 11(23), 3871.
\item Al-Breiki, H., Rehman, M. H. U., Salah, K., \& Svetinovic, D. (2020).
Trustworthy blockchain oracles: review, comparison, and open research
challenges. IEEE Access, 8, 85675-85685.
\item Alladi, T., Chamola, V., Rodrigues, J. J., \& Kozlov, S. A. (2019).
Blockchain in smart grids: A review on different use cases. Sensors,
19(22), 4862.
\item Alexander, C. (2005). The present and future of financial risk management.
Journal of Financial Econometrics, 3(1), 3-25.
\item Andersen, T. G., Bollerslev, T., Christoffersen, P. F., \& Diebold,
F. X. (2013). Financial risk measurement for financial risk management.
In Handbook of the Economics of Finance (Vol. 2, pp. 1127-1220). Elsevier.
\item Andryukhin, A. A. (2019, March). Phishing attacks and preventions
in blockchain based projects. In 2019 international conference on
engineering technologies and computer science (EnT) (pp. 15-19). IEEE.
\item Ante, L., Fiedler, I., \& Strehle, E. (2021). The influence of stablecoin
issuances on cryptocurrency markets. Finance Research Letters, 41,
101867.
\item Aravindhan, K., \& Karthiga, R. R. (2013). One time password: A survey.
International Journal of Emerging Trends in Engineering and Development,
1(3), 613-623.
\item Arshadi, N. (2019). Application of Blockchain Protocol to Wealth Management.
The Journal of Wealth Management, 21(4), 122-129.
\item Azeez, N. A., Misra, S., Margaret, I. A., \& Fernandez-Sanz, L. (2021).
Adopting automated whitelist approach for detecting phishing attacks.
Computers \& Security, 108, 102328.
\item Bacis, E., Facchinetti, D., Guarnieri, M., Rosa, M., Rossi, M., \&
Paraboschi, S. (2021, August). I told you tomorrow: Practical time-locked
secrets using smart contracts. In Proceedings of the 16th International
Conference on Availability, Reliability and Security (pp. 1-10). 
\item Balci, O. (1995, December). Principles and techniques of simulation
validation, verification, and testing. In Proceedings of the 27th
conference on Winter simulation (pp. 147-154).
\item Baliga, A., Subhod, I., Kamat, P., \& Chatterjee, S. (2018). Performance
evaluation of the quorum blockchain platform. arXiv preprint arXiv:1809.03421.
\item Banerjee, M., Lee, J., Chen, Q., \& Choo, K. K. R. (2018, July). Blockchain-based
security layer for identification and isolation of malicious things
in IoT: A conceptual design. In 2018 27th International Conference
on Computer Communication and Networks (ICCCN) (pp. 1-6). IEEE.
\item Barkadehi, M. H., Nilashi, M., Ibrahim, O., Fardi, A. Z., \& Samad,
S. (2018). Authentication systems: A literature review and classification.
Telematics and Informatics, 35(5), 1491-1511.
\item Bartoletti, M., Carta, S., Cimoli, T., \& Saia, R. (2020). Dissecting
Ponzi schemes on Ethereum: identification, analysis, and impact. Future
Generation Computer Systems, 102, 259-277.
\item Bartoletti, M., Chiang, J. H. Y., \& Lafuente, A. L. (2021). SoK:
lending pools in decentralized finance. In Financial Cryptography
and Data Security. FC 2021 International Workshops: CoDecFin, DeFi,
VOTING, and WTSC, Virtual Event, March 5, 2021, Revised Selected Papers
25 (pp. 553-578). Springer Berlin Heidelberg.
\item Belchior, R., Vasconcelos, A., Guerreiro, S., \& Correia, M. (2021).
A survey on blockchain interoperability: Past, present, and future
trends. ACM Computing Surveys (CSUR), 54(8), 1-41.
\item Bellare, M., \& Neven, G. (2007, February). Identity-based multi-signatures
from RSA. In Cryptographers’ Track at the RSA Conference (pp. 145-162).
Berlin, Heidelberg: Springer Berlin Heidelberg.
\item Beniiche, A. (2020). A study of blockchain oracles. arXiv preprint
arXiv:2004.07140.
\item Bernstein, D. J., \& Lange, T. (2017). Post-quantum cryptography.
Nature, 549(7671), 188-194.
\item Bertsimas, D., \& Lo, A. W. (1998). Optimal control of execution costs.
Journal of financial markets, 1(1), 1-50.
\item Bilali, G. (2011). Know your customer-or not. U. Tol. L. Rev., 43,
319.
\item Boehm, B. W. (1983). Seven basic principles of software engineering.
Journal of Systems and Software, 3(1), 3-24.
\item Bosu, A., Iqbal, A., Shahriyar, R., \& Chakraborty, P. (2019). Understanding
the motivations, challenges and needs of blockchain software developers:
A survey. Empirical Software Engineering, 24(4), 2636-2673.
\item Briola, A., Vidal-Tomás, D., Wang, Y., \& Aste, T. (2023). Anatomy
of a Stablecoin’s failure: The Terra-Luna case. Finance Research Letters,
51, 103358.
\item Brophy, R. (2020). Blockchain and insurance: a review for operations
and regulation. Journal of financial regulation and compliance, 28(2),
215-234.
\item Brown, D. J., \& Werner, J. (1995). Arbitrage and existence of equilibrium
in infinite asset markets. The Review of Economic Studies, 62(1),
101-114.
\item Brown, M., Trautmann, S. T., \& Vlahu, R. (2017). Understanding bank-run
contagion. Management Science, 63(7), 2272-2282.
\item Bumblauskas, D., Mann, A., Dugan, B., \& Rittmer, J. (2020). A blockchain
use case in food distribution: Do you know where your food has been?.
International Journal of Information Management, 52, 102008.
\item Cai, C. W. (2018). Disruption of financial intermediation by FinTech:
a review on crowdfunding and blockchain. Accounting \& Finance, 58(4),
965-992.
\item Caldarelli, G. (2020). Understanding the blockchain oracle problem:
A call for action. Information, 11(11), 509.
\item Caldarelli, G., \& Ellul, J. (2021). The blockchain oracle problem
in decentralized finance—a multivocal approach. Applied Sciences,
11(16), 7572.
\item Caldarelli, G. (2022). Overview of blockchain oracle research. Future
Internet, 14(6), 175.
\item Calvo, G. (2012). Financial crises and liquidity shocks a bank-run
perspective. European Economic Review, 56(3), 317-326.
\item Caldarelli, G., \& Ellul, J. (2021). The blockchain oracle problem
in decentralized finance—a multivocal approach. Applied Sciences,
11(16), 7572.
\item Chakraborty, P., Shahriyar, R., Iqbal, A., \& Bosu, A. (2018, October).
Understanding the software development practices of blockchain projects:
a survey. In Proceedings of the 12th ACM/IEEE international symposium
on empirical software engineering and measurement (pp. 1-10).
\item Chaliasos, S., Charalambous, M. A., Zhou, L., Galanopoulou, R., Gervais,
A., Mitropoulos, D., \& Livshits, B. (2023). Smart contract and defi
security: Insights from tool evaluations and practitioner surveys.
arXiv preprint arXiv:2304.02981.
\item Chen, T., Zhang, Y., Li, Z., Luo, X., Wang, T., Cao, R., ... \& Zhang,
X. (2019, November). Tokenscope: Automatically detecting inconsistent
behaviors of cryptocurrency tokens in ethereum. In Proceedings of
the 2019 ACM SIGSAC conference on computer and communications security
(pp. 1503-1520).
\item Chen, T. H. (2020). Do you know your customer? Bank risk assessment
based on machine learning. Applied Soft Computing, 86, 105779.
\item Chen, W., Guo, X., Chen, Z., Zheng, Z., \& Lu, Y. (2020, July). Phishing
Scam Detection on Ethereum: Towards Financial Security for Blockchain
Ecosystem. In IJCAI (Vol. 7, pp. 4456-4462).
\item Chen, J., Xia, X., Lo, D., Grundy, J., \& Yang, X. (2021). Maintenance-related
concerns for post-deployed Ethereum smart contract development: issues,
techniques, and future challenges. Empirical Software Engineering,
26(6), 117.
\item Chen, J., Lin, D., \& Wu, J. (2022). Do cryptocurrency exchanges fake
trading volumes? An empirical analysis of wash trading based on data
mining. Physica A: Statistical Mechanics and its Applications, 586,
126405. 
\item Chen, Y. L., Chang, Y. T., \& Yang, J. J. (2023). Cryptocurrency hacking
incidents and the price dynamics of Bitcoin spot and futures. Finance
Research Letters, 103955.
\item Chia, V., Hartel, P., Hum, Q., Ma, S., Piliouras, G., Reijsbergen,
D., ... \& Szalachowski, P. (2018, July). Rethinking blockchain security:
Position paper. In 2018 IEEE International Conference on Internet
of Things (iThings) and IEEE Green Computing and Communications (GreenCom)
and IEEE Cyber, Physical and Social Computing (CPSCom) and IEEE Smart
Data (SmartData) (pp. 1273-1280). IEEE.
\item Chiew, K. L., Yong, K. S. C., \& Tan, C. L. (2018). A survey of phishing
attacks: Their types, vectors and technical approaches. Expert Systems
with Applications, 106, 1-20.
\item Corbet, S., Cumming, D. J., Lucey, B. M., Peat, M., \& Vigne, S. A.
(2020). The destabilising effects of cryptocurrency cybercriminality.
Economics Letters, 191, 108741.
\item Cousaert, S., Xu, J., \& Matsui, T. (2022, May). Sok: Yield aggregators
in defi. In 2022 IEEE International Conference on Blockchain and Cryptocurrency
(ICBC) (pp. 1-14). IEEE.
\item De Bondt, W. F. (1998). A portrait of the individual investor. European
economic review, 42(3-5), 831-844.
\item Deng, T., Fu, J., Zheng, Q., Wu, J., \& Pi, Y. (2019). Performance-based
wind-resistant optimization design for tall building structures. Journal
of Structural Engineering, 145(10), 04019103.
\item Denning, P. J. (2005). Is computer science science?. Communications
of the ACM, 48(4), 27-31.
\item Desikan, S., \& Ramesh, G. (2006). Software testing: principles and
practice. Pearson Education India.
\item Desolda, G., Ferro, L. S., Marrella, A., Catarci, T., \& Costabile,
M. F. (2021). Human factors in phishing attacks: a systematic literature
review. ACM Computing Surveys (CSUR), 54(8), 1-35.
\item Dijk, O. (2017). Bank run psychology. Journal of Economic Behavior
\& Organization, 144, 87-96.
\item Dimitri, N. (2022). Consensus: Proof of Work, Proof of Stake and structural
alternatives. Enabling the Internet of Value: How Blockchain Connects
Global Businesses, 29-36.
\item Dos Santos, S., Singh, J., Thulasiram, R. K., Kamali, S., Sirico,
L., \& Loud, L. (2022, June). A new era of blockchain-powered decentralized
finance (DeFi)-a review. In 2022 IEEE 46th Annual Computers, Software,
and Applications Conference (COMPSAC) (pp. 1286-1292). IEEE.
\item Edwards, F. R. (1999). Hedge funds and the collapse of long-term capital
management. Journal of Economic Perspectives, 13(2), 189-210.
\item Elton, E. J., Gruber, M. J., Brown, S. J., \& Goetzmann, W. N. (2009).
Modern portfolio theory and investment analysis. John Wiley \& Sons.
\item Engler, D., Chen, D. Y., Hallem, S., Chou, A., \& Chelf, B. (2001).
Bugs as deviant behavior: A general approach to inferring errors in
systems code. ACM SIGOPS Operating Systems Review, 35(5), 57-72.
\item Erdem, E., \& Sandıkkaya, M. T. (2018). OTPaaS—One time password as
a service. IEEE Transactions on Information Forensics and Security,
14(3), 743-756.
\item Farooq, M. S., Kalim, Z., Qureshi, J. N., Rasheed, S., \& Abid, A.
(2022). A blockchain-based framework for distributed agile software
development. IEEE Access, 10, 17977-17995.
\item Feng, C., Li, N., Wong, M. H., \& Zhang, M. (2019). Initial coin offerings,
blockchain technology, and white paper disclosures. Mingyue, Initial
Coin Offerings, Blockchain Technology, and White Paper Disclosures
(March 25, 2019).
\item Fraga-Lamas, P., \& Fernandez-Carames, T. M. (2020). Fake news, disinformation,
and deepfakes: Leveraging distributed ledger technologies and blockchain
to combat digital deception and counterfeit reality. IT professional,
22(2), 53-59.
\item Fuertes, A. M., Muradoglu, G., \& Ozturkkal, B. (2014). A behavioral
analysis of investor diversification. The European Journal of Finance,
20(6), 499-523.
\item Gao, B., Wang, H., Xia, P., Wu, S., Zhou, Y., Luo, X., \& Tyson, G.
(2020). Tracking counterfeit cryptocurrency end-to-end. Proceedings
of the ACM on Measurement and Analysis of Computing Systems, 4(3),
1-28.
\item Gatteschi, V., Lamberti, F., \& Demartini, C. (2020). Blockchain technology
use cases. Advanced applications of Blockchain technology, 91-114.
\item Girasa, R. (2018). Regulation of cryptocurrencies and blockchain technologies:
national and international perspectives. Springer.
\item Goel, A. K., Bisht, V. S., \& Chaudhary, S. (2023, June). Multisignature
Crypto Wallet Paper. In 2023 8th International Conference on Communication
and Electronics Systems (ICCES) (pp. 476-479). IEEE.
\item Gonzalez, L. (2020). Blockchain, herding and trust in peer-to-peer
lending. Managerial Finance, 46(6), 815-831.
\item Goyal, V., Abraham, A., Sanyal, S., \& Han, S. Y. (2005, April). The
N/R one time password system. In International Conference on Information
Technology: Coding and Computing (ITCC'05)-Volume II (Vol. 1, pp.
733-738). IEEE.
\item Grassi, L., Lanfranchi, D., Faes, A., \& Renga, F. M. (2022). Do we
still need financial intermediation? The case of decentralized finance–DeFi.
Qualitative Research in Accounting \& Management, 19(3), 323-347.
\item Green, R., \& Ledgard, H. (2011). Coding guidelines: Finding the art
in the science. Communications of the ACM, 54(12), 57-63.
\item Greener, I. (2006). Nick Leeson and the collapse of Barings Bank:
Socio-technical networks and the ‘Rogue Trader’. Organization, 13(3),
421-441.
\item Grobys, K. (2021). When the blockchain does not block: on hackings
and uncertainty in the cryptocurrency market. Quantitative Finance,
21(8), 1267-1279.
\item Groza, B., \& Petrica, D. (2005). One-time passwords for uncertain
number of authentications. Proceedings of CSCS15.
\item Gu, M. (2010). Wind-resistant studies on tall buildings and structures.
Science China Technological Sciences, 53, 2630-2646.
\item Guo, Y., \& Liang, C. (2016). Blockchain application and outlook in
the banking industry. Financial innovation, 2, 1-12.
\item Guo, H., \& Yu, X. (2022). A survey on blockchain technology and its
security. Blockchain: research and applications, 3(2), 100067. 
\item Hafid, A., Hafid, A. S., \& Samih, M. (2020). Scaling blockchains:
A comprehensive survey. IEEE access, 8, 125244-125262.
\item Halgamuge, M. N. (2022). Estimation of the success probability of
a malicious attacker on blockchain-based edge network. Computer Networks,
219, 109402.
\item Haller, N., Metz, C., Nesser, P., \& Straw, M. (1998). A one-time
password system (No. rfc2289).
\item Hammi, B., Zeadally, S., Adja, Y. C. E., Del Giudice, M., \& Nebhen,
J. (2021). Blockchain-based solution for detecting and preventing
fake check scams. IEEE Transactions on Engineering Management, 69(6),
3710-3725.
\item Han, J., Song, M., Eom, H., \& Son, Y. (2021, March). An efficient
multi-signature wallet in blockchain using bloom filter. In Proceedings
of the 36th Annual ACM Symposium on Applied Computing (pp. 273-281).
\item Hanson, R. (2007). Logarithmic market scoring rules for modular combinatorial
information aggrega- tion. The Journal of Prediction Markets, 1(1),
3-15.
\item Harvey, C. R., Ramachandran, A., \& Santoro, J. (2021). DeFi and the
Future of Finance. John Wiley \& Sons.
\item Hassija, V., Bansal, G., Chamola, V., Kumar, N., \& Guizani, M. (2020).
Secure lending: Blockchain and prospect theory-based decentralized
credit scoring model. IEEE Transactions on Network Science and Engineering,
7(4), 2566-2575.
\item He, S., Wu, Q., Luo, X., Liang, Z., Li, D., Feng, H., ... \& Li, Y.
(2018). A social-network-based cryptocurrency wallet-management scheme.
IEEE Access, 6, 7654-7663.
\item Herremans, D., \& Low, K. W. (2022). Forecasting Bitcoin volatility
spikes from whale transactions and CryptoQuant data using Synthesizer
Transformer models. arXiv preprint arXiv:2211.08281.
\item Holzmann, G. J. (2002). The logic of bugs. ACM SIGSOFT Software Engineering
Notes, 27(6), 81-87.
\item Horcher, K. A. (2011). Essentials of financial risk management. John
Wiley \& Sons.
\item Huang, F., \& Bin, L. I. U. (2017). Software defect prevention based
on human error theories. Chinese Journal of Aeronautics, 30(3), 1054-1070.
\item Huang, F., Liu, B., \& Huang, B. (2012, July). A taxonomy system to
identify human error causes for software defects. In The 18th international
conference on reliability and quality in design (pp. 44-49). 
\item Huang, Y., Wang, H., Wu, L., Tyson, G., Luo, X., Zhang, R., ... \&
Jiang, X. (2020). Characterizing eosio blockchain. arXiv preprint
arXiv:2002.05369. 
\item Houy, S., Schmid, P., \& Bartel, A. (2023). Security Aspects of Cryptocurrency
Wallets—A Systematic Literature Review. ACM Computing Surveys, 56(1),
1-31.
\item Huang, C. Y., Ma, S. P., \& Chen, K. T. (2011). Using one-time passwords
to prevent password phishing attacks. Journal of Network and Computer
Applications, 34(4), 1292-1301.
\item Hunhevicz, J. J., \& Hall, D. M. (2020). Do you need a blockchain
in construction? Use case categories and decision framework for DLT
design options. Advanced Engineering Informatics, 45, 101094.
\item Jain, A. K., \& Gupta, B. B. (2016). A novel approach to protect against
phishing attacks at client side using auto-updated white-list. EURASIP
Journal on Information Security, 2016, 1-11.
\item Jakobsson, M., \& Juels, A. (1999, September). Proofs of work and
bread pudding protocols. In Secure Information Networks: Communications
and Multimedia Security IFIP TC6/TC11 Joint Working Conference on
Communications and Multimedia Security (CMS’99) September 20–21, 1999,
Leuven, Belgium (pp. 258-272). Boston, MA: Springer US.
\item Jansen, B. J. (1998). The graphical user interface. ACM SIGCHI Bulletin,
30(2), 22-26.
\item Jorion, P., \& Khoury, S. (1996). Financial risk management. Cambridge/Massachusetts.
\item Jorion, P. (2000). Risk management lessons from long‐term capital
management. European financial management, 6(3), 277-300.
\item Karamitsos, I., Papadaki, M., \& Al Barghuthi, N. B. (2018). Design
of the blockchain smart contract: A use case for real estate. Journal
of Information Security, 9(3), 177-190.
\item Karantias, K. (2020). Sok: A taxonomy of cryptocurrency wallets. Cryptology
ePrint Archive.
\item Karastilo, R. (2020). David vs Goliath (You against the Markets),
A dynamic programming approach to separate the impact and timing of
trading costs. Physica A: Statistical Mechanics and its Applications,
545 (May 2020), 122848.
\item Karbeer, B. (2016). Fighting Uncertainty with Uncertainty. Available
at SSRN 2715424.
\item Kareem, Q. (2021-I). Trade Execution: To Trade or Not To Trade. Working
Paper.
\item Kareem, Q. (2021-II). The Democratization of Wealth Management: Hedged
Mutual Fund Blockchain Protocol. Working Paper.
\item Kasaliya, A. (2021). Do Traders Become Rogues or Do Rogues Become
Traders? The Om of Jerome and the Karma of Kerviel. Corp. \& Bus.
LJ, 2, 88.
\item Kasaliya, A. (2022). Bringing Risk Parity To The DeFi Party: A Complete
Solution To The Crypto Asset Management Conundrum. Initial Draft.
\item Kashyap, R. (2023). Arguably Adequate Aqueduct Algorithm: Crossing
A Bridge-Less Block-Chain Chasm. Finance Research Letters, 58, 104421.
2. 
\item Kashyap, R. (2024). Risk Management: A Slow Walk On A Tight Rope.
Journal of Investing, Forthcoming. 
\item Kaur, G., Habibi Lashkari, A., Sharafaldin, I., \& Habibi Lashkari,
Z. (2023). Smart Contracts and DeFi Security and Threats. In Understanding
Cybersecurity Management in Decentralized Finance: Challenges, Strategies,
and Trends (pp. 91-111). Cham: Springer International Publishing.
\item Kaushik, A., Choudhary, A., Ektare, C., Thomas, D., \& Akram, S. (2017,
May). Blockchain—literature survey. In 2017 2nd IEEE International
Conference on Recent Trends in Electronics, Information \& Communication
Technology (RTEICT) (pp. 2145-2148). IEEE.
\item Kereiakes, Evan, Marco Di Maggio Do Kwon, and Nicholas Platias. \textquotedbl Terra
money: Stability and adoption.\textquotedbl{} White Paper, Apr (2019).
\item Khan, S. N., Loukil, F., Ghedira-Guegan, C., Benkhelifa, E., \& Bani-Hani,
A. (2021). Blockchain smart contracts: Applications, challenges, and
future trends. Peer-to-peer Networking and Applications, 14, 2901-2925.
\item Kiong, L. V. (2021). How to Maximize Return in DeFi: A Beginner’s
Guide to Yield Farming and Liquidity Mining. Liew Voon Kiong.
\item Kirda, E., \& Kruegel, C. (2006). Protecting users against phishing
attacks. The Computer Journal, 49(5), 554-561. 
\item Knuth, D. E. (2014). Art of computer programming, volume 2: Seminumerical
algorithms. Addison-Wesley Professional.
\item Krawiec, K. D. (2000). Accounting for greed: Unraveling the rogue
trader mystery. Or. L. Rev., 79, 301. 
\item Krawiec, K. D. (2009). The return of the rogue. Ariz. L. Rev., 51,
127.
\item Kuo, T. T., Zavaleta Rojas, H., \& Ohno-Machado, L. (2019). Comparison
of blockchain platforms: a systematic review and healthcare examples.
Journal of the American Medical Informatics Association, 26(5), 462-478.
\item Ladleif, J., \& Weske, M. (2020, October). Time in blockchain-based
process execution. In 2020 IEEE 24th International Enterprise Distributed
Object Computing Conference (EDOC) (pp. 217-226). IEEE.
\item Lai, W. J., Hsueh, C. W., \& Wu, J. L. (2019, July). A fully decentralized
time-lock encryption system on blockchain. In 2019 IEEE International
Conference on Blockchain (Blockchain) (pp. 302-307). IEEE.
\item Lamport, L. (1981). Password authentication with insecure communication.
Communications of the ACM, 24(11), 770-772.
\item Laurent, M., Kaaniche, N., Le, C., \& Vander Plaetse, M. (2018, July).
A blockchain-based access control scheme. In SECRYPT 2018: 15th International
Conference on Security and Cryptography (Vol. 2, pp. 168-176). Scitepress.
\item Lee, S., Lee, J., \& Lee, Y. (2022). Dissecting the Terra-LUNA crash:
Evidence from the spillover effect and information flow. Finance Research
Letters, 103590.
\item Li, Y., Duan, R. B., Li, Q. S., Li, Y. G., \& Huang, X. (2020, October).
Wind-resistant optimal design of tall buildings based on improved
genetic algorithm. In Structures (Vol. 27, pp. 2182-2191). Elsevier. 
\item Li, W., Bu, J., Li, X., Peng, H., Niu, Y., \& Chen, X. (2022). A Survey
of DeFi Security: Challenges and Opportunities. arXiv preprint arXiv:2206.11821.
\item Li, W., Bu, J., Li, X., \& Chen, X. (2022, August). Security analysis
of DeFi: Vulnerabilities, attacks and advances. In 2022 IEEE International
Conference on Blockchain (Blockchain) (pp. 488-493). IEEE.
\item Li, Y., Liu, H., \& Tan, Y. (2022, May). POLYBRIDGE: A Crosschain
Bridge for Heterogeneous Blockchains. In 2022 IEEE International Conference
on Blockchain and Cryptocurrency (ICBC) (pp. 1-2). IEEE.
\item Liu, Y., Xiong, Z., Hu, Q., Niyato, D., Zhang, J., Miao, C., ... \&
Tian, Z. (2022). VRepChain: A decentralized and privacy-preserving
reputation system for social Internet of Vehicles based on blockchain.
IEEE Transactions on Vehicular Technology, 71(12), 13242-13253.
\item Liu, Y., Yu, W., Ai, Z., Xu, G., Zhao, L., \& Tian, Z. (2022). A blockchain-empowered
federated learning in healthcare-based cyber physical systems. IEEE
Transactions on Network Science and Engineering.
\item Liu, Y., Zhang, C., Yan, Y., Zhou, X., Tian, Z., \& Zhang, J. (2022).
A semi-centralized trust management model based on blockchain for
data exchange in iot system. IEEE Transactions on Services Computing,
16(2), 858-871.
\item Lo, S. K., Xu, X., Staples, M., \& Yao, L. (2020). Reliability analysis
for blockchain oracles. Computers \& Electrical Engineering, 83, 106582.
\item Lowenstein, R. (2001). When genius failed: The rise and fall of Long-Term
Capital Management. Random House trade paperbacks.
\item Lu, Y. (2019). The blockchain: State-of-the-art and research challenges.
Journal of Industrial Information Integration, 15, 80-90.
\item Lyons, R. K., \& Viswanath-Natraj, G. (2023). What keeps stablecoins
stable?. Journal of International Money and Finance, 131, 102777.
\item Lys, L., \& Potop-Butucaru, M. (2022, May). Distributed Blockchain
Price Oracle. In International Conference on Networked Systems (pp.
37-51). Cham: Springer International Publishing.
\item MacKenzie, D. (2003). Long-Term Capital Management and the sociology
of arbitrage. Economy and society, 32(3), 349-380.
\item Mammadzada, K., Iqbal, M., Milani, F., García-Bañuelos, L., \& Matulevičius,
R. (2020). Blockchain oracles: a framework for blockchain-based applications.
In Business Process Management: Blockchain and Robotic Process Automation
Forum: BPM 2020 Blockchain and RPA Forum, Seville, Spain, September
13–18, 2020, Proceedings 18 (pp. 19-34). Springer International Publishing.
\item Malhotra, D., Saini, P., \& Singh, A. K. (2022). How blockchain can
automate KYC: systematic review. Wireless Personal Communications,
122(2), 1987-2021.
\item Manahov, V. (2022). Cryptocurrency liquidity during extreme price
movements: is there a problem with virtual money?. In Commodities
(pp. 731-762). Chapman and Hall/CRC.
\item Marchesi, M., Marchesi, L., \& Tonelli, R. (2018, October). An agile
software engineering method to design blockchain applications. In
Proceedings of the 14th Central and Eastern European Software Engineering
Conference Russia (pp. 1-8).
\item McBee, M. P., \& Wilcox, C. (2020). Blockchain technology: principles
and applications in medical imaging. Journal of digital imaging, 33,
726-734.
\item Meng, M., Steinhardt, S., \& Schubert, A. (2018). Application programming
interface documentation: What do software developers want?. Journal
of Technical Writing and Communication, 48(3), 295-330.
\item Merkle, R. C. (1990). A fast software one-way hash function. Journal
of Cryptology, 3, 43-58.
\item Miraz, M. H., Excell, P. S., \& Rafiq, M. K. S. B. (2021). Evaluation
of green alternatives for blockchain proof-of-work (PoW) approach.
Annals of Emerging Technologies in Computing (AETiC), 54-59.
\item Mohan, V. (2022). Automated market makers and decentralized exchanges:
a DeFi primer. Financial Innovation, 8(1), 20.
\item Mohanta, B. K., Panda, S. S., \& Jena, D. (2018, July). An overview
of smart contract and use cases in blockchain technology. In 2018
9th international conference on computing, communication and networking
technologies (ICCCNT) (pp. 1-4). IEEE.
\item Mohanty, S. K., \& Tripathy, S. (2022). Siovchain: time-lock contract
based privacy-preserving data sharing in siov. IEEE Transactions on
Intelligent Transportation Systems, 23(12), 24071-24082.
\item Moore, T., \& Christin, N. (2013). Beware the middleman: Empirical
analysis of Bitcoin-exchange risk. In Financial Cryptography and Data
Security: 17th International Conference, FC 2013, Okinawa, Japan,
April 1-5, 2013, Revised Selected Papers 17 (pp. 25-33). Springer
Berlin Heidelberg.
\item Moro-Visconti, R., \& Cesaretti, A. (2023). The Cryptocurrency Crash
of 2022: Which Lessons for the Future?. In Digital Token Valuation:
Cryptocurrencies, NFTs, Decentralized Finance, and Blockchains (pp.
395-410). Cham: Springer Nature Switzerland.
\item M'Raihi, D., Machani, S., Pei, M., \& Rydell, J. (2011). Totp: Time-based
one-time password algorithm (No. rfc6238).
\item Nagase, T., Hisatoku, T., \& Yamazaki, S. (1993, April). Wind resistant
design and response control of tall building. In Structural Engineering
in Natural Hazards Mitigation (pp. 532-537). ASCE.
\item Nakashima, T., Oyama, M., Hisada, H., \& Ishii, N. (1999). Analysis
of software bug causes and its prevention. Information and Software
technology, 41(15), 1059-1068.
\item Naor, M., \& Yung, M. (1989, February). Universal one-way hash functions
and their cryptographic applications. In Proceedings of the twenty-first
annual ACM symposium on Theory of computing (pp. 33-43).
\item Norman, D. A. (1983). Design rules based on analyses of human error.
Communications of the ACM, 26(4), 254-258. 
\item Ofoeda, J., Boateng, R., \& Effah, J. (2019). Application programming
interface (API) research: A review of the past to inform the future.
International Journal of Enterprise Information Systems (IJEIS), 15(3),
76-95.
\item Oosthoek, K. (2021). Flash crash for cash: Cyber threats in decentralized
finance. arXiv preprint arXiv:2106.10740.
\item Ostern, N. K., \& Riedel, J. (2021). Know-your-customer (KYC) requirements
for initial coin offerings. Business \& Information Systems Engineering,
63(5), 551-567.
\item Oulasvirta, A., Dayama, N. R., Shiripour, M., John, M., \& Karrenbauer,
A. (2020). Combinatorial optimization of graphical user interface
designs. Proceedings of the IEEE, 108(3), 434-464.
\item Ouriemmi, O., \& Gérard, B. (2023). Control dynamics in rogue trading:
Sovereignty and exception-to-the-rule attitudes in the contemporary
financial sphere. Critical Perspectives on Accounting, 91, 102414.
\item Pal, O., Alam, B., Thakur, V., \& Singh, S. (2021). Key management
for blockchain technology. ICT express, 7(1), 76-80.
\item Pasdar, A., Dong, Z., \& Lee, Y. C. (2021). Blockchain oracle design
patterns. arXiv preprint arXiv:2106.09349. 
\item Pasdar, A., Lee, Y. C., \& Dong, Z. (2023). Connect API with blockchain:
A survey on blockchain oracle implementation. ACM Computing Surveys,
55(10), 1-39.
\item Patel, S. B., Bhattacharya, P., Tanwar, S., \& Kumar, N. (2020). Kirti:
A blockchain-based credit recommender system for financial institutions.
IEEE Transactions on Network Science and Engineering, 8(2), 1044-1054.
\item Perold, A. F. (1988). The implementation shortfall: Paper versus reality.
Journal of Portfolio Management, 14(3), 4.
\item Peterson, M. (2018). Blockchain and the future of financial services.
The Journal of Wealth Management, 21(1), 124-131.
\item Pierro, G. A., Rocha, H., Tonelli, R., \& Ducasse, S. (2020, February).
Are the gas prices oracle reliable? a case study using the ethgasstation.
In 2020 IEEE International Workshop on Blockchain Oriented Software
Engineering (IWBOSE) (pp. 1-8). IEEE.
\item Pierro, G. A., Rocha, H., Ducasse, S., Marchesi, M., \& Tonelli, R.
(2022). A user-oriented model for oracles’ gas price prediction. Future
Generation Computer Systems, 128, 142-157.
\item Pillai, A., Saraswat, V., \& VR, A. (2019). Smart wallets on blockchain—attacks
and their costs. In Smart City and Informatization: 7th International
Conference, iSCI 2019, Guangzhou, China, November 12–15, 2019, Proceedings
7 (pp. 649-660). Springer Singapore.
\item Pokhrel, S. R., \& Choi, J. (2020). Federated learning with blockchain
for autonomous vehicles: Analysis and design challenges. IEEE Transactions
on Communications, 68(8), 4734-4746.
\item Prewett, K. W., Prescott, G. L., \& Phillips, K. (2020). Blockchain
adoption is inevitable—Barriers and risks remain. Journal of Corporate
accounting \& finance, 31(2), 21-28.
\item Purkait, S. (2012). Phishing counter measures and their effectiveness–literature
review. Information Management \& Computer Security, 20(5), 382-420.
\item Puthal, D., Malik, N., Mohanty, S. P., Kougianos, E., \& Das, G. (2018).
Everything you wanted to know about the blockchain: Its promise, components,
processes, and problems. IEEE Consumer Electronics Magazine, 7(4),
6-14.
\item Qayyum, A., Qadir, J., Janjua, M. U., \& Sher, F. (2019). Using blockchain
to rein in the new post-truth world and check the spread of fake news.
IT Professional, 21(4), 16-24.
\item Qin, K., Zhou, L., Livshits, B., \& Gervais, A. (2021, October). Attacking
the defi ecosystem with flash loans for fun and profit. In Financial
Cryptography and Data Security: 25th International Conference, FC
2021, Virtual Event, March 1–5, 2021, Revised Selected Papers, Part
I (pp. 3-32). Berlin, Heidelberg: Springer Berlin Heidelberg.
\item Qiu, H., Wu, X., Zhang, S., Leung, V. C., \& Cai, W. (2019, December).
ChainIDE: A cloud-based integrated development environment for cross-blockchain
smart contracts. In 2019 IEEE International Conference on Cloud Computing
Technology and Science (CloudCom) (pp. 317-319). IEEE.
\item Rabieinejad, E., Yazdinejad, A., Parizi, R. M., \& Dehghantanha, A.
(2023). Generative adversarial networks for cyber threat hunting in
ethereum blockchain. Distributed Ledger Technologies: Research and
Practice.
\item Rasmussen, J. (1997). Risk management in a dynamic society: a modelling
problem. Safety science, 27(2-3), 183-213.
\item Ray, P. P. (2017). An introduction to dew computing: definition, concept
and implications. IEEE Access, 6, 723-737.
\item Reinhart, C. M., \& Rogoff, K. S. (2009). This time is different:
Eight centuries of financial folly. princeton university press.
\item Saleh, F. (2021). Blockchain without waste: Proof-of-stake. The Review
of financial studies, 34(3), 1156-1190.
\item Sargent, R. G. (2010, December). Verification and validation of simulation
models. In Proceedings of the 2010 winter simulation conference (pp.
166-183). IEEE.
\item Sayed, R. H. (2019). Potential of blockchain technology to solve fake
diploma problem.
\item Schär, F. (2021). Decentralized finance: On blockchain-and smart contract-based
financial markets. FRB of St. Louis Review.
\item Shanaev, S., Sharma, S., Ghimire, B., \& Shuraeva, A. (2020). Taming
the blockchain beast? Regulatory implications for the cryptocurrency
Market. Research in International Business and Finance, 51, 101080.
\item Shleifer, A., \& Vishny, R. W. (1997). The limits of arbitrage. The
Journal of Finance, 52(1), 35-55.
\item Simiu, E., \& Scanlan, R. H. (1996). Wind effects on structures: fundamentals
and applications to design (Vol. 688). New York: John Wiley.
\item Slamka, C., Skiera, B., \& Spann, M. (2012). Prediction market performance
and market liquidity: A comparison of automated market makers. IEEE
Transactions on Engineering Management, 60(1), 169-185.
\item Stephen, R., \& Alex, A. (2018, August). A review on blockchain security.
In IOP conference series: materials science and engineering (Vol.
396, No. 1, p. 012030). IOP Publishing.
\item Stone, D. (2021). Trustless, privacy-preserving blockchain bridges.
arXiv preprint arXiv:2102.04660.
\item Suratkar, S., Shirole, M., \& Bhirud, S. (2020, September). Cryptocurrency
wallet: A review. In 2020 4th international conference on computer,
communication and signal processing (ICCCSP) (pp. 1-7). IEEE.
\item Swan, M. (2016). Blockchain temporality: Smart contract time specifiability
with blocktime. In Rule Technologies. Research, Tools, and Applications:
10th International Symposium, RuleML 2016, Stony Brook, NY, USA, July
6-9, 2016. Proceedings 10 (pp. 184-196). Springer International Publishing.
\item Thakkar, P., Nathan, S., \& Viswanathan, B. (2018, September). Performance
benchmarking and optimizing hyperledger fabric blockchain platform.
In 2018 IEEE 26th international symposium on modeling, analysis, and
simulation of computer and telecommunication systems (MASCOTS) (pp.
264-276). IEEE.
\item Thomas, P., \& Michael, G. (2007). How to Cheat at VoIP Security.
Editora Syngress, Sao Paulo, 5.
\item Tian, Z., Li, M., Qiu, M., Sun, Y., \& Su, S. (2019). Block-DEF: A
secure digital evidence framework using blockchain. Information Sciences,
491, 151-165.
\item Torre, N. G., \& Rudd, A. (2004). The portfolio management problem
of individual investors: A quantitative perspective. The Journal of
Wealth Management, 7(1), 56-63.
\item Trozze, A., Kleinberg, B., \& Davies, T. (2021). Detecting DeFi Securities
Violations from Token Smart Contract Code with Random Forest Classification.
arXiv preprint arXiv:2112.02731.
\item Trozze, A., Kamps, J., Akartuna, E. A., Hetzel, F. J., Kleinberg,
B., Davies, T., \& Johnson, S. D. (2022). Cryptocurrencies and future
financial crime. Crime Science, 11, 1-35.
\item Uhlig, H. (2010). A model of a systemic bank run. Journal of Monetary
Economics, 57(1), 78-96.
\item Uhlig, H. (2022). A Luna-tic Stablecoin Crash (No. w30256). National
Bureau of Economic Research.
\item Vacca, A., Di Sorbo, A., Visaggio, C. A., \& Canfora, G. (2021). A
systematic literature review of blockchain and smart contract development:
Techniques, tools, and open challenges. Journal of Systems and Software,
174, 110891.
\item Varshney, G., Misra, M., \& Atrey, P. K. (2016). A survey and classification
of web phishing detection schemes. Security and Communication Networks,
9(18), 6266-6284.
\item Wang, S., Yuan, Y., Wang, X., Li, J., Qin, R., \& Wang, F. Y. (2018).
An overview of smart contract: architecture, applications, and future
trends. In 2018 IEEE Intelligent Vehicles Symposium (IV) (pp. 108-113).
IEEE.
\item Wang, D., Wu, S., Lin, Z., Wu, L., Yuan, X., Zhou, Y., ... \& Ren,
K. (2021, May). Towards a first step to understand flash loan and
its applications in defi ecosystem. In Proceedings of the Ninth International
Workshop on Security in Blockchain and Cloud Computing (pp. 23-28).
\item Wang, B., Liu, H., Liu, C., Yang, Z., Ren, Q., Zheng, H., \& Lei,
H. (2021, May). Blockeye: Hunting for DeFi attacks on blockchain.
In 2021 IEEE/ACM 43rd International Conference on Software Engineering:
Companion Proceedings (ICSE-Companion) (pp. 17-20). IEEE.
\item Wang, B., Yuan, X., Duan, L., Ma, H., Su, C., \& Wang, W. (2022).
DeFiScanner: Spotting DeFi Attacks Exploiting Logic Vulnerabilities
on Blockchain. IEEE Transactions on Computational Social Systems.
\item Weber, I., Gramoli, V., Ponomarev, A., Staples, M., Holz, R., Tran,
A. B., \& Rimba, P. (2017, September). On availability for blockchain-based
systems. In 2017 IEEE 36th Symposium on Reliable Distributed Systems
(SRDS) (pp. 64-73). IEEE.
\item Wendl, M., Doan, M. H., \& Sassen, R. (2023). The environmental impact
of cryptocurrencies using proof of work and proof of stake consensus
algorithms: A systematic review. Journal of Environmental Management,
326, 116530.
\item Werapun, W., Karode, T., Arpornthip, T., Suaboot, J., Sangiamkul,
E., \& Boonrat, P. (2022, December). The Flash Loan Attack Analysis
(FAA) Framework—A Case Study of the Warp Finance Exploitation. In
Informatics (Vol. 10, No. 1, p. 3). MDPI.
\item Werner, S. M., Perez, D., Gudgeon, L., Klages-Mundt, A., Harz, D.,
\& Knottenbelt, W. J. (2021). Sok: Decentralized finance (defi). arXiv
preprint arXiv:2101.08778. 
\item Whitaker, A. (2019). Art and blockchain: A primer, history, and taxonomy
of blockchain use cases in the arts. Artivate, 8(2), 21-46.
\item Wu, J., Yuan, Q., Lin, D., You, W., Chen, W., Chen, C., \& Zheng,
Z. (2020). Who are the phishers? phishing scam detection on ethereum
via network embedding. IEEE Transactions on Systems, Man, and Cybernetics:
Systems, 52(2), 1156-1166.
\item Wu, S., Wang, D., He, J., Zhou, Y., Wu, L., Yuan, X., ... \& Ren,
K. (2021). Defiranger: Detecting price manipulation attacks on defi
applications. arXiv preprint arXiv:2104.15068.
\item Xia, P., Wang, H., Zhang, B., Ji, R., Gao, B., Wu, L., ... \& Xu,
G. (2020). Characterizing cryptocurrency exchange scams. Computers
\& Security, 98, 101993.
\item Xu, J., Paruch, K., Cousaert, S., \& Feng, Y. (2021). Sok: Decentralized
exchanges (dex) with automated market maker (amm) protocols. ACM Computing
Surveys.
\item Xu, J., \& Feng, Y. (2022). Reap the Harvest on Blockchain: A Survey
of Yield Farming Protocols. IEEE Transactions on Network and Service
Management.
\item Xue, Y., Fu, J., Su, S., Bhuiyan, Z. A., Qiu, J., Lu, H., ... \& Tian,
Z. (2022, July). Preventing Price Manipulation Attack by Front-Running.
In Advances in Artificial Intelligence and Security: 8th International
Conference on Artificial Intelligence and Security, ICAIS 2022, Qinghai,
China, July 15–20, 2022, Proceedings, Part III (pp. 309-322). Cham:
Springer International Publishing.
\item Yadav, S. P., Agrawal, K. K., Bhati, B. S., Al-Turjman, F., \& Mostarda,
L. (2022). Blockchain-based cryptocurrency regulation: An overview.
Computational Economics, 59(4), 1659-1675. 
\item Yeoh, P. (2017). Regulatory issues in blockchain technology. Journal
of Financial Regulation and Compliance, 25(2), 196-208. 
\item Yeung, K. (2019). Regulation by blockchain: the emerging battle for
supremacy between the code of law and code as law. The Modern Law
Review, 82(2), 207-239.
\item Yilmaz, M., Tasel, S., Tuzun, E., Gulec, U., O’Connor, R. V., \& Clarke,
P. M. (2019). Applying blockchain to improve the integrity of the
software development process. In Systems, Software and Services Process
Improvement: 26th European Conference, EuroSPI 2019, Edinburgh, UK,
September 18–20, 2019, Proceedings 26 (pp. 260-271). Springer International
Publishing.
\item Zaimi, R., Hafidi, M., \& Lamia, M. (2020, December). Survey paper:
Taxonomy of website anti-phishing solutions. In 2020 Seventh International
Conference on Social Networks Analysis, Management and Security (SNAMS)
(pp. 1-8). IEEE.
\item Zamani, E., He, Y., \& Phillips, M. (2020). On the security risks
of the blockchain. Journal of Computer Information Systems, 60(6),
495-506.
\item Zeng, X., Hao, N., Zheng, J., \& Xu, X. (2019). A consortium blockchain
paradigm on hyperledger-based peer-to-peer lending system. China Communications,
16(8), 38-50.
\item Zetzsche, D. A., Arner, D. W., \& Buckley, R. P. (2020). Decentralized
finance. Journal of Financial Regulation, 6(2), 172-203.
\item Zhang, P., Schmidt, D. C., White, J., \& Lenz, G. (2018). Blockchain
technology use cases in healthcare. In Advances in computers (Vol.
111, pp. 1-41). Elsevier.
\item Zhang, Y., Xu, C., Cheng, N., Li, H., Yang, H., \& Shen, X. (2019).
Chronos \$\textasciicircum\{\{\textbackslash mathbf+\}\} \$+: An
Accurate Blockchain-Based Time-Stamping Scheme for Cloud Storage.
IEEE Transactions on Services Computing, 13(2), 216-229.
\item Zhang, H., Zou, X., Xie, G., \& Li, Z. (2022, December). Blockchain
Multi-signature Wallet System. In Blockchain Technology and Application:
5th CCF China Blockchain Conference, CBCC 2022, Wuxi, China, December
23–25, 2022, Proceedings (p. 31). Springer Nature.
\item Zheng, Z., Xie, S., Dai, H. N., Chen, W., Chen, X., Weng, J., \& Imran,
M. (2020). An overview on smart contracts: Challenges, advances and
platforms. Future Generation Computer Systems, 105, 475-491.
\item Zheng, W., Liu, B., Dai, H. N., Jiang, Z., Zheng, Z., \& Imran, M.
(2022). Unravelling token ecosystem of eosio blockchain. arXiv preprint
arXiv:2202.11201.
\item Zhou, Q., Huang, H., Zheng, Z., \& Bian, J. (2020). Solutions to scalability
of blockchain: A survey. Ieee Access, 8, 16440-16455.
\item Zhou, L., Xiong, X., Ernstberger, J., Chaliasos, S., Wang, Z., Wang,
Y., ... \& Gervais, A. (2022). Sok: Decentralized finance (defi) attacks.
Cryptology ePrint Archive.
\item Zhu, J., Khan, A., \& Akcora, C. G. (2023). Core-based Trend Detection
in Blockchain Networks. arXiv preprint arXiv:2303.14241.
\item Zou, W., Lo, D., Kochhar, P. S., Le, X. B. D., Xia, X., Feng, Y.,
... \& Xu, B. (2019). Smart contract development: Challenges and opportunities.
IEEE Transactions on Software Engineering, 47(10), 2084-2106.
\end{enumerate}
\pagebreak{}

\part{\label{part:Appendix-of-Supplementary}Appendix of Supplementary
Material}

\section{\label{sec:End-notes}Appendix: End-notes and Explanations}
\begin{enumerate}
\item \label{enu:Blockchain-Terms}The following terms are important to
understand how blockchain operates: The Ledger; Linked Time stamping;
Merkle Trees; Byzantine fault tolerance; Proof Of Work.
\begin{enumerate}
\item A blockchain is a distributed ledger with growing lists of records
(blocks) that are securely linked together via cryptographic hashes.
Each block contains a cryptographic hash of the previous block, a
timestamp, and transaction data (generally represented as a Merkle
tree, where data nodes are represented by leaves). Since each block
contains information about the previous block, they effectively form
a chain (compare linked list data structure), with each additional
block linking to the ones before it. Consequently, blockchain transactions
are irreversible in that, once they are recorded, the data in any
given block cannot be altered retroactively without altering all subsequent
blocks. \href{https://en.wikipedia.org/wiki/Blockchain}{Blockchain,  Wikipedia Link}
\item In cryptography and computer science, a hash tree or Merkle tree is
a tree in which every \textquotedbl leaf\textquotedbl{} (node) is
labelled with the cryptographic hash of a data block, and every node
that is not a leaf (called a branch, inner node, or inode) is labelled
with the cryptographic hash of the labels of its child nodes. A hash
tree allows efficient and secure verification of the contents of a
large data structure. A hash tree is a generalization of a hash list
and a hash chain. \href{https://en.wikipedia.org/wiki/Merkle_tree}{Merkle Tree,  Wikipedia Link}
\item Although blockchain records are not unalterable, since blockchain
forks are possible, blockchains may be considered secure by design
and exemplify a distributed computing system with high Byzantine fault
tolerance. 
\item A Byzantine fault (also Byzantine generals problem, interactive consistency,
source congruency, error avalanche, Byzantine agreement problem, and
Byzantine failure) is a condition of a computer system, particularly
distributed computing systems, where components may fail and there
is imperfect information on whether a component has failed. The term
takes its name from an allegory, the \textquotedbl Byzantine generals
problem\textquotedbl , developed to describe a situation in which,
in order to avoid catastrophic failure of the system, the system's
actors must agree on a concerted strategy, but some of these actors
are unreliable. \href{https://en.wikipedia.org/wiki/Byzantine_fault}{Byzantine Fault,  Wikipedia Link}
\item Proof of work (PoW) is a form of cryptographic proof in which one
party (the prover) proves to others (the verifiers) that a certain
amount of a specific computational effort has been expended. Verifiers
can subsequently confirm this expenditure with minimal effort on their
part (Jakobsson \& Juels 1999). The purpose of proof-of-work algorithms
is not proving that certain work was carried out or that a computational
puzzle was \textquotedbl solved\textquotedbl , but deterring manipulation
of data by establishing large energy and hardware-control requirements
to be able to do so. \href{https://en.wikipedia.org/wiki/Proof_of_work}{Proof of Work,  Wikipedia Link}
\item Proof-of-work systems have been criticized by environmentalists for
their energy consumption. Several alternatives are being developed
due to the environment concerns of to PoW algorithms (Miraz et al.,
2021; Dimitri 2022). 
\item Proof-of-stake (PoS) protocols are a class of consensus mechanisms
for blockchains that work by selecting validators in proportion to
their quantity of holdings in the associated cryptocurrency (Saleh
2021; Wendl et al., 2023). This is done to avoid the computational
cost of proof-of-work (POW) schemes. The first functioning use of
PoS for cryptocurrency was Peer-coin in 2012, although the scheme,
on the surface, still resembled a POW. \href{https://en.wikipedia.org/wiki/Proof_of_stake}{Proof of Stake,  Wikipedia Link}
\end{enumerate}
\item \label{enu:Federated-learning-(also}Federated learning (also known
as collaborative learning) is a machine learning technique that trains
an algorithm via multiple independent sessions, each using its own
dataset. This approach stands in contrast to traditional centralized
machine learning techniques where local datasets are merged into one
training session, as well as to approaches that assume that local
data samples are identically distributed.

Federated learning enables multiple actors to build a common, robust
machine learning model without sharing data, thus addressing critical
issues such as data privacy, data security, data access rights and
access to heterogeneous data. \href{https://en.wikipedia.org/wiki/Federated_learning}{Federated Learning,  Wikipedia Link}
\item \label{enu:Types-Yield-Enhancement-Services}The following are the
four main types of blockchain yield enhancement services. We can also
consider them as the main types of financial products available in
decentralized finance: 
\begin{enumerate}
\item \label{enu:Single-sided-staking-allows}Single-Sided Staking: This
allows users to earn yield by providing liquidity for one type of
asset, in contrast to liquidity provisioning on AMMs, which requires
a pair of assets. \href{https://docs.saucerswap.finance/features/single-sided-staking}{Single Sided Staking,  SuacerSwap Link}
\begin{enumerate}
\item Bancor is an example of a provider who supports single sided staking.
Bancor natively supports Single-Sided Liquidity Provision of tokens
in a liquidity pool. This is one of the main benefits to liquidity
providers that distinguishes Bancor from other DeFi staking protocols.
Typical AMM liquidity pools require a liquidity provider to provide
two assets. Meaning, if you wish to deposit \textquotedbl TKN1\textquotedbl{}
into a pool, you would be forced to sell 50\% of that token and trade
it for \textquotedbl TKN2\textquotedbl . When providing liquidity,
your deposit is composed of both TKN1 and TKN2 in the pool. Bancor
Single-Side Staking changes this and enables liquidity providers to:
Provide only the token they hold (TKN1 from the example above) Collect
liquidity providers fees in TKN1. \href{https://docs.bancor.network/about-bancor-network/faqs/single-side-liquidity}{Single Sided Staking,  Bancor Link}
\end{enumerate}
\item \label{enu:AMM-Liquidity-Pairs}AMM Liquidity Pairs (AMM LP): A constant-function
market maker (CFMM) is a market maker with the property that that
the amount of any asset held in its inventory is completely described
by a well-defined function of the amounts of the other assets in its
inventory (Hanson 2007). \href{https://en.wikipedia.org/wiki/Constant_function_market_maker}{Constant Function Market Maker,  Wikipedia Link}

This is the most common type of market maker liquidity pool. Other
types of market makers are discussed in Mohan (2022). All of them
can be grouped under the category Automated Market Makers. Hence the
name AMM Liquidity Pairs. A more general discussion of AMMs, without
being restricted only to the blockchain environment, is given in (Slamka,
Skiera \& Spann 2012).
\item \label{enu:LP-Token-Staking:}LP Token Staking: LP staking is a valuable
way to incentivize token holders to provide liquidity. When a token
holder provides liquidity as mentioned earlier in Point (\ref{enu:AMM-Liquidity-Pairs})
they receive LP tokens. LP staking allows the liquidity providers
to stake their LP tokens and receive project tokens tokens as rewards.
This mitigates the risk of impermanent loss and compensates for the
loss. \href{https://defactor.com/liquidity-provider-staking-introduction-guide/}{Liquidity Provider Staking,  DeFactor Link}
\begin{enumerate}
\item Note that this is also a type of single sided staking discussed in
Point (\ref{enu:Single-sided-staking-allows}). The key point to remember
is that the LP Tokens can be considered as receipts for the crypto
assets deposits in an AMM LP Point (\ref{enu:AMM-Liquidity-Pairs}).
These LP Token receipts can be further staked to generate additional
yield.
\end{enumerate}
\item \label{enu:Lending:-Crypto-lending}Lending: Crypto lending is the
process of depositing cryptocurrency that is lent out to borrowers
in return for regular interest payments. Payments are typically made
in the form of the cryptocurrency that is deposited and can be compounded
on a daily, weekly, or monthly basis. \href{https://www.investopedia.com/crypto-lending-5443191}{Crypto Lending,  Investopedia Link};
\href{https://defiprime.com/decentralized-lending}{DeFi Lending,  DeFiPrime Link};
\href{https://crypto.com/price/categories/lending}{Top Lending Coins by Market Capitalization,  Crypto.com Link}.
\begin{enumerate}
\item Crypto lending is very common on decentralized finance projects and
also in centralized exchanges. Centralized cryptocurrency exchanges
are online platforms used to buy and sell cryptocurrencies. They are
the most common means that investors use to buy and sell cryptocurrency
holdings. \href{https://www.investopedia.com/tech/what-are-centralized-cryptocurrency-exchanges/}{Centralized Cryptocurrency Exchanges,  Investopedia Link}
\item Lending is a very active area of research both on blockchain and off
chain (traditional finance) as well (Cai 2018; Zeng et al., 2019;
Bartoletti, Chiang \& Lafuente 2021; Gonzalez 2020; Hassija et al.,
2020; Patel et al. , 2020). 
\end{enumerate}
\end{enumerate}
\item \label{enu:Decentralized-finance}Decentralized finance (often stylized
as DeFi) offers financial instruments without relying on intermediaries
such as brokerages, exchanges, or banks by using smart contracts on
a blockchain. \href{https://en.wikipedia.org/wiki/Decentralized_finance}{Decentralized Finance (DeFi), Wikipedia Link}
\item \label{enu:Decentralized-exchanges-(DEX)}Decentralized exchanges
(DEX) are a type of cryptocurrency exchange which allows for direct
peer-to-peer cryptocurrency transactions to take place without the
need for an intermediary, so they are fundamentally differerent from
centralized exchanges (CEX). \href{https://en.wikipedia.org/wiki/Decentralized_finance\#Decentralized_exchanges}{Decentralized Exchanges (DEX),  Wikipedia Link}
\item \label{enu:Centralized-cryptocurrency-excha}Centralized cryptocurrency
exchanges (CEX), or just centralized exchanges, act as an intermediary
between a buyer and a seller and make money through commissions and
transaction fees. You can imagine a CEX to be similar to a stock exchange
but for digital assets. \href{https://corporatefinanceinstitute.com/resources/cryptocurrency/cryptocurrency-exchanges/}{Centralized Cryptocurrency Exchanges (CEX),  Wikipedia Link};
\href{https://www.investopedia.com/tech/what-are-centralized-cryptocurrency-exchanges/}{Centralized Cryptocurrency Exchanges (CEX),  Investopedia Link};
\href{https://www.coindesk.com/learn/what-is-a-cex-centralized-exchanges-explained/}{Centralized Cryptocurrency Exchanges (CEX),  CoinDesk Link};
\href{https://www.coindesk.com/learn/centralized-exchange-cex-vs-decentralized-exchange-dex-whats-the-difference/}{CEX vs DEX Difference,  CoinDesk Link}
\item \label{enu:An-application-programming-interface}An application programming
interface (API) is in contrast with a User Interface. It is a way
for two or more computer programs to communicate with each other.
It is a type of software interface, offering a service to other pieces
of software. \href{https://en.wikipedia.org/wiki/API}{Application Programming Interface (API),  Wikipedia Link}
\item \label{enu:A-smart-contract}A smart contract is a computer program
or a transaction protocol that is intended to automatically execute,
control or document events and actions according to the terms of a
contract or an agreement. The objectives of smart contracts are the
reduction of need for trusted intermediators, arbitration costs, and
fraud losses, as well as the reduction of malicious and accidental
exceptions (Wang et al., 2018; Mohanta et al., 2018; Zou et al., 2019;
Zheng et al., 2020). \href{https://en.wikipedia.org/wiki/Smart_contract}{Smart Contract,  Wikipedia Link}
\item \label{enu:A-one-time-password}A one-time password (OTP), also known
as a one-time PIN, one-time authorization code (OTAC) or dynamic password,
is a password that is valid for only one login session or transaction,
on a computer system or other digital device. \href{https://en.wikipedia.org/wiki/One-time_password}{One Time Password (OTP),  Wikipedia Link}
\item \label{enu:Public-key-cryptography,-or}Public-key cryptography, or
asymmetric cryptography, is the field of cryptographic systems that
use pairs of related keys. Each key pair consists of a public key
and a corresponding private key. Key pairs are generated with cryptographic
algorithms based on mathematical problems termed one-way functions.
Security of public-key cryptography depends on keeping the private
key secret; the public key can be openly distributed without compromising
security. \href{https://en.wikipedia.org/wiki/Public-key_cryptography}{Public Key Cryptography,  Wikipedia Link}
\item \label{enu:One-Way-Function}In computer science, a one-way function
is a function that is easy to compute on every input, but hard to
invert given the image of a random input. Here, \textquotedbl easy\textquotedbl{}
and \textquotedbl hard\textquotedbl{} are to be understood in the
sense of computational complexity theory, specifically the theory
of polynomial time problems. \href{https://en.wikipedia.org/wiki/One-way_function}{One-Way Function,  Wikipedia Link}
\item \label{enu:A-one-way-hash}A one-way hash function, also known as
a message digest, is a mathematical function that takes a variable-length
input string and converts it into a fixed-length binary sequence that
is computationally difficult to invert - that is, generate the original
string from the hash (Thomas \& Michael 2007). 
\begin{enumerate}
\item A one-way hash function is a function $F$ which accepts an arbitrarily
large input $x$, and produces a small fixed-size output $y$. That
is, $y=F(x).$ Further, no other input $x'$ can be found (although
many such inputs almost certainly exist) which will generate $y$.
Because of this, a small $y$ can authenticate an arbitrarily large
$x$. This property is crucial for the convenient authentication of
large amounts of information (Merkle 1990).
\end{enumerate}
\item \label{enu:Cloud-computing}Cloud computing is the on-demand availability
of computer system resources, especially data storage (cloud storage)
and computing power, without direct active management by the user.
Large clouds often have functions distributed over multiple locations,
each of which is a data center. Cloud computing relies on sharing
of resources to achieve coherence. \href{https://en.wikipedia.org/wiki/One-way_function}{One-Way Function,  Wikipedia Link}
\item \label{enu:KYC}The know your customer or know your client (KYC) guidelines
in financial services require that professionals make an effort to
verify the identity, suitability, and risks involved with maintaining
a business relationship. \href{https://en.wikipedia.org/wiki/Know_your_customer}{Know Your Customer,  Wikipedia Link};
\href{https://www.investopedia.com/terms/k/knowyourclient.asp}{Know Your Client,  Investopedia Link}
\item \label{enu:The-graphical-user}The graphical user interface, or GUI,
is a form of user interface that allows users to interact with electronic
devices through graphical icons and audio indicators such as primary
notation, instead of text-based UIs, typed command labels or text
navigation. GUIs were introduced in reaction to the perceived steep
learning curve of command-line interfaces (CLIs), which require commands
to be typed on a computer keyboard. \href{https://en.wikipedia.org/wiki/Graphical_user_interface}{Graphical User Interface,  Wikipedia Link}
\begin{enumerate}
\item Web-based user interfaces or web user interfaces (WUI) that accept
input and provide output by generating web pages viewed by the user
using a web browser program. \href{https://en.wikipedia.org/wiki/User_interface}{User Interface,  Wikipedia Link}
\end{enumerate}
\item \label{enu:Stablecoin}A Stablecoin is a type of cryptocurrency where
the value of the digital asset is supposed to be pegged to a reference
asset, which is either fiat money, exchange-traded commodities (such
as precious metals or industrial metals), or another cryptocurrency.
\href{https://en.wikipedia.org/wiki/Stablecoin}{Stable Coin,  Wikipedia Link}
\item \label{enu:A-cryptocurrency-wallet}A cryptocurrency wallet is a device,
physical medium, program or a service which stores the public and/or
private keys{[}3{]} for cryptocurrency transactions. In addition to
this basic function of storing the keys, a cryptocurrency wallet more
often offers the functionality of encrypting and/or signing information.
Signing can for example result in executing a smart contract, a cryptocurrency
transaction (see \textquotedbl bitcoin transaction\textquotedbl{}
image), identification, or legally signing a 'document'. \href{https://en.wikipedia.org/wiki/Cryptocurrency_wallet}{Cryptocurrency Wallet,  Wikipedia Link}
\item \label{enu:Terra-May-2022}In May 2022, the Terra blockchain was temporarily
halted after the collapse of the stablecoin TerraUSD (UST) and Luna,
in an event that wiped out almost \$45 billion in market capitalisation
within a week.
\begin{enumerate}
\item Terra is a blockchain protocol and payment platform used for algorithmic
stablecoins. The project was created in 2018 by Terraform Labs, a
startup co-founded by Do Kwon and Daniel Shin. It is most known for
its Terra stablecoin and the associated Luna reserve asset cryptocurrency.
\href{https://en.wikipedia.org/wiki/Terra_(blockchain)}{Terra (Blockchain),  Wikipedia Link}
\end{enumerate}
\item \label{enu:Long-Term-Capital-Management}Long-Term Capital Management
L.P. (LTCM) was a highly leveraged hedge fund. In 1998, it received
a \$3.6 billion bailout from a group of 14 banks, in a deal brokered
and put together by the Federal Reserve Bank of New York. \href{https://en.wikipedia.org/wiki/Long-Term_Capital_Management}{Long-Term Capital Management,  Wikipedia Link}
\begin{enumerate}
\item LTCM was initially successful, with annualized returns (after fees)
of around 21\% in its first year, 43\% in its second year and 41\%
in its third year. However, in 1998 it lost \$4.6 billion in less
than four months due to a combination of high leverage and exposure
to the 1997 Asian financial crisis and 1998 Russian financial crisis.
The fund was liquidated and dissolved in early 2000.
\end{enumerate}
\begin{doublespace}
\item \label{Arbitrage, Wikipedia Link}In economics and finance, arbitrage
is the practice of taking advantage of a price difference between
two or more markets: striking a combination of matching deals that
capitalize upon the imbalance, the profit being the difference between
the market prices. When used by academics, an arbitrage is a (imagined,
hypothetical, thought experiment) transaction that involves no negative
cash flow at any probabilistic or temporal state and a positive cash
flow in at least one state; in simple terms, it is the possibility
of a risk-free profit after transaction costs. For example, an arbitrage
opportunity is present when there is the opportunity to instantaneously
buy something for a low price and sell it for a higher price. \href{https://en.wikipedia.org/wiki/Arbitrage}{Arbitrage, Wikipedia Link}
\end{doublespace}
\item \label{enu:Software-Testing-Validation}We would like to highlight
the following points to help with the actual coding of the software
(Boehm 1983; Balci 1995; Denning 2005; Desikan \& Ramesh 2006; Sargent
2010; Green \& Ledgard 2011; Knuth 2014). The algorithm we have provided
acts mostly as detailed implementation guidelines. Many cases and
error conditions need to be handled appropriately during implementation.
Alternate implementation simplifications, time conventions, and counters
are possible and can be accommodated accordingly. There might even
be some issues - or bugs - with the variables, counters and timing.
These are due to limitations of not actually testing scenarios using
a full fledged software system. But the gist of what we have provided
should carry over to the coding stage with very little changes. Conditional
statements such as - if ... then ... else - can be used depending
on the implementation language and other efficiency considerations
as necessary.
\item \label{enu:An-integrated-development-environment}An integrated development
environment (IDE) is a software application that provides comprehensive
facilities for software development. An IDE normally consists of at
least a source-code editor, build automation tools, and a debugger.
\href{https://en.wikipedia.org/wiki/Integrated_development_environment}{Integrated Development Environment, Wikipedia Link};
\href{https://ethereum.org/en/developers/docs/ides/}{Integrated Development Environments (IDEs),  Ethereum. Org Link};
\href{https://www.alchemy.com/overviews/solidity-ide}{The 7 Best Solidity IDEs for Developers (2023),  Alchemy Website}
\item \label{enu:Distributed-agile-software}Distributed agile software
development is a research area that considers the effects of applying
the principles of agile software development to a globally distributed
development setting, with the goal of overcoming challenges in projects
which are geographically distributed. \href{https://en.wikipedia.org/wiki/Distributed_agile_software_development}{Distributed Agile Software Development,  Wikipedia Link}
\begin{enumerate}
\item In software development, agile practices (sometimes written \textquotedbl Agile\textquotedbl )
include requirements discovery and solutions improvement through the
collaborative effort of self-organizing and cross-functional teams
with their customer(s)/end user(s). \href{https://en.wikipedia.org/wiki/Agile_software_development}{Agile Software Development,  Wikipedia Link}
\end{enumerate}
\item \label{enu:Blockchain-Bridges}Blockchain bridges work just like the
bridges we know in the physical world. Just as a physical bridge connects
two physical locations, a blockchain bridge connects two blockchain
ecosystems. Bridges facilitate communication between blockchains through
the transfer of information and assets. \href{https://ethereum.org/en/bridges/}{Blockchain Bridges,  Ethereum.Org Website Link};
\href{https://hacken.io/discover/blockchain-bridges/}{Blockchain Bridges 101,  hacken.io Website Link}
\item \label{enu:Ethereum-is-a}Ethereum is a decentralized blockchain with
smart contract functionality. Ether (Abbreviation: ETH) is the native
cryptocurrency of the platform. Among cryptocurrencies, ether is second
only to bitcoin in market capitalization. It is open-source software.
\href{https://en.wikipedia.org/wiki/Ethereum}{Ethereum,  Wikipedia Link}
\item \label{enu:Phishing-is-a}Phishing is a form of social engineering
and scam where attackers deceive people into revealing sensitive information
or installing malware such as ransomware. \href{https://en.wikipedia.org/wiki/Phishing}{Phishing,  Wikipedia Link} 
\begin{enumerate}
\item The term \textquotedbl phishing\textquotedbl{} was first recorded
in 1995 in the cracking toolkit AOHell, but may have been used earlier
in the hacker magazine 2600. It is a variation of fishing and refers
to the use of lures to \textquotedbl fish\textquotedbl{} for sensitive
information.
\end{enumerate}
\item \label{enu:Multi-Sign Wallet}In contrast to simple cryptocurrency
wallets requiring just one party to sign a transaction, multi-sig
wallets require multiple parties to sign a transaction.{[}19{]} Multisignature
wallets are designed for increased security. Usually, a multisignature
algorithm produces a joint signature that is more compact than a collection
of distinct signatures from all users. \href{https://en.wikipedia.org/wiki/Cryptocurrency_wallet\#Multisignature_wallet}{Multi-Signature Wallet,  Wikipedia Link}
\item \label{enu:A-whitelist-is}A whitelist is a list or register of entities
that are being provided a particular privilege, service, mobility,
access or recognition. Entities on the list will be accepted, approved
and/or recognized. Whitelisting is the reverse of blacklisting, the
practice of identifying entities that are denied, unrecognized, or
ostracized. \href{https://en.wikipedia.org/wiki/Whitelist}{Whitelist,  Wikipedia Link}
\begin{enumerate}
\item Whitelists are mostly used in the context of Initial Coin Offerings
(ICOs) or in terms of withdrawal addresses. It is a list of cryptocurrency
addresses that are deemed trustworthy. Only addresses that appear
on such lists can withdraw funds from exchange accounts. \href{https://coinmarketcap.com/academy/glossary/whitelist}{Whitelist,  Coin Market Cap Link}
\end{enumerate}
\item \label{enu:An-initial-coin}An initial coin offering (ICO) or initial
currency offering is a type of funding using cryptocurrencies. It
is often a form of crowdfunding, although a private ICO which does
not seek public investment is also possible. In an ICO, a quantity
of cryptocurrency is sold in the form of \textquotedbl tokens\textquotedbl{}
(\textquotedbl coins\textquotedbl ) to speculators or investors,
in exchange for legal tender or other (generally established and more
stable) cryptocurrencies such as Bitcoin or Ether. The tokens are
promoted as future functional units of currency if or when the ICO's
funding goal is met and the project successfully launches. \href{https://en.wikipedia.org/wiki/Initial_coin_offering}{Initial Coin Offering,  Wikipedia Link}

Any cryptocurrency or blockchain company looking to raise funds to
create an app, service or new coin can use an ICO to raise funds.
It is widely seen as the cryptocurrency world’s answer to initial
public offerings (IPOs) — and they were especially popular during
the crypto bubble of 2017. \href{https://coinmarketcap.com/academy/glossary/initial-coin-offering-ico}{Initial Coin Offering (ICO),  Coin Market Cap Link}
\item \label{enu:A-timelock-is}A timelock is a piece of code that locks
a certain functionality of a smart contract until a specific amount
of time has passed. Most often, this is the ability to transfer a
token out of the contract. \href{https://www.certik.com/resources/blog/Timelock}{Timelock,  Certik Link}

At its core, a timelock is an additional piece of code that restricts
functionality within a smart contract to a specific window of time.
The simplest form could look something like this simple “if” statement:
``if (block.timestamp < \_timelockTime) \{ revert ErrorNotReady(block.timestamp,
\_timelockTime); \}'' \href{https://blog.chain.link/timelock-smart-contracts/}{Timelock Smart Contracts,   Chain-Link Website}
\item \label{enu:A-bank-run}A bank run or run on the bank occurs when many
clients withdraw their money from a bank, because they believe the
bank may fail in the near future. \href{https://en.wikipedia.org/wiki/Bank_run}{Bank Run,  Wikipedia Link}
\item \label{enu:A-cryptocurrency-whale,}A cryptocurrency whale, more commonly
known as a \textquotedbl crypto whale\textquotedbl{} or just a \textquotedbl whale,\textquotedbl{}
is a cryptocurrency community term that refers to individuals or entities
that hold large amounts of cryptocurrency. Whales own enough cryptocurrency
to influence currency markets. \href{https://www.investopedia.com/terms/b/bitcoin-whale.asp}{Crypto Whale,  Investopedia Link}
\item \label{enu:Implementation-Shortfall}In financial markets, implementation
shortfall is the difference between the decision price and the final
execution price (including commissions, taxes, etc.) for a trade.
This is also known as the \textquotedbl slippage\textquotedbl .
\href{https://en.wikipedia.org/wiki/Implementation_shortfall}{Implementation Shortfall,  Wikipedia Link}
\item \label{enu:The-block-time}The block time is the average time it takes
for the network to generate one extra block in the blockchain. By
the time of block completion, the included data becomes verifiable.
In cryptocurrency, this is practically when the transaction takes
place, so a shorter block time means faster transactions. The block
time for Ethereum is set to between 14 and 15 seconds, while for bitcoin
it is on average 10 minutes. \href{https://en.wikipedia.org/wiki/Blockchain\#Block_time}{Block Time,  Wikipedia Link}
\item \label{enu:Blockchain-oracles-are}Blockchain oracles are entities
that connect blockchains to external systems, thereby enabling smart
contracts to execute based upon inputs and outputs from the real world.
\href{https://chain.link/education/blockchain-oracles}{Blockchain Oracles,  Chain-Link Link}

A blockchain oracle is a third-party service that connects smart contracts
with the outside world, primarily to feed information in from the
world, but also the reverse. \href{https://en.wikipedia.org/wiki/Blockchain_oracle}{Blockchain Oracle,  Wikipedia Link}
\item \label{enu:Finance-AUM}In finance, assets under management (AUM),
sometimes called fund under management, measures the total market
value of all the financial assets which an individual or financial
institution—such as a mutual fund, venture capital firm, or depository
institution—or a decentralized network protocol controls, typically
on behalf of a client. \href{https://en.wikipedia.org/wiki/Assets_under_management}{Assets Under Management, Wikipedia Link}
\item \label{enu:TVL}In decentralized finance, Total value locked represents
the number of assets that are currently being staked in a specific
protocol.\href{https://coinmarketcap.com/alexandria/glossary/total-value-locked-tvl}{Total Value Locked, CoinMarketCap Link}
\item \label{enu:Uniswap-is-a}Uniswap is a decentralized cryptocurrency
exchange that uses a set of smart contracts (liquidity pools) to execute
trades on its exchange. \href{https://en.wikipedia.org/wiki/Uniswap}{Uniswap, Wikipedia Link}
\item \label{enu:PancakeSwap-(CAKE)-is}PancakeSwap (CAKE) is the leading
decentralized exchange on BNB Chain.\href{https://coinmarketcap.com/alexandria/article/what-is-pancakeswap}{Pancake Swap, CoinMarketCap Link}
\item \label{enu:Nansen-is-a}Nansen is a blockchain analytics platform
that enriches on-chain data with millions of wallet labels. Crypto
investors use Nansen to discover opportunities, perform due diligence
and defend their portfolios with our real-time dashboards and alerts.
Nansen (\href{https://www.nansen.ai/about}{Nansen,  Website Link}.)
was previously known as Apeboard (\href{https://apeboard.finance/dashboard}{Apeboard Link}).
\item \label{enu:DefiLlama-is-the}DeFiLlama is the largest TVL aggregator
for DeFi (Decentralized Finance). \href{https://defillama.com}{DeFiLlama, Website Link}
\item \label{enu:CoinMarketCap}CoinMarketCap is a leading price-tracking
website for crypto-assets in the cryptocurrency space. Its mission
is to make crypto discoverable and efficient globally by empowering
retail users with unbiased, high quality and accurate information
for drawing their own informed conclusions. It was founded in May
2013 by Brandon Chez. \href{https://coinmarketcap.com/about/}{CoinMarketCap, Website Link}
\item \label{enu:Crypto-Ranking}A ranking of cryptocurrencies, including
symbols for the various tokens, by market capitalization is available
on the CoinMarketCap website. We are using the data as of May-25-2022,
when the first version of this article was written. \href{https://coinmarketcap.com}{CoinMarketCap Cryptocurrency Ranking,  Website Link}
\end{enumerate}

\end{document}